\documentclass[journal]{IEEEtran}
\usepackage{stmaryrd}
\usepackage{amsfonts}
\usepackage{amsmath}
\usepackage{amssymb}
\usepackage{cite}
\usepackage{graphicx}
\usepackage{subfigure}
\usepackage{amsthm}
\usepackage{algorithmic}
\usepackage{algorithm}
\usepackage{stfloats}
\usepackage{textcomp}

\newtheorem{Theorem}{Theorem}

\newtheorem{Lemma}[Theorem]{Lemma}

\newtheorem{Algorithm}{Algorithm}

\def\QED{\IEEEQED\vspace{0.1in}}

\begin{document}

\title{The Synthesis and Analysis of Stochastic Switching Circuits}

\author{Hongchao~Zhou,
        Po-Ling~Loh,
        Jehoshua~Bruck,~\IEEEmembership{Fellow,~IEEE}
\thanks{This work was supported in part by the NSF Expeditions in Computing Program under grant CCF-0832824. This paper was presented in part at IEEE International Symposium on Information Theory (ISIT), Seoul, Korea, June 2009.}
\thanks{H. Zhou and J. Bruck are with the Department
of Electrical Engineering, California Institute of Technology, Pasadena, CA, 91125.
{\it Email: hzhou@caltech.edu, bruck@caltech.edu}}
\thanks{P. Loh is with the Department of Statistics, University of California, Berkeley, CA 94720. {\it Email: ploh@berkeley.edu}}
}
\maketitle

\begin{abstract}
  Stochastic switching circuits are relay circuits that consist of
  stochastic switches called pswitches. The study of
  stochastic switching circuits has widespread applications in many
  fields of computer science, neuroscience, and biochemistry. In this
  paper, we discuss several properties of stochastic switching
  circuits, including robustness, expressibility, and probability approximation.

  First, we study the robustness, namely, the effect caused by introducing an error of size
  $\epsilon$ to each pswitch in a stochastic circuit. We analyze two
  constructions and prove that simple series-parallel circuits are robust
  to small error perturbations, while general series-parallel circuits
  are not. Specifically, the total error introduced by perturbations
  of size less than $\epsilon$ is bounded by a constant multiple of
  $\epsilon$ in a simple series-parallel circuit, independent of the
  size of the circuit.

  Next, we study the expressibility of stochastic switching circuits: Given an integer $q$ and a pswitch set
  $S=\{\frac{1}{q},\frac{2}{q},\dots,\frac{q-1}{q}\}$, can we
  synthesize any rational probability with denominator $q^n$ (for
  arbitrary $n$) with a simple series-parallel stochastic switching
  circuit? We generalize previous results and prove that when $q$ is a
  multiple of $2$ or $3$, the answer is yes. We also show that when
  $q$ is a prime number larger than $3$, the answer is no.

  Probability approximation is studied for a general case of an arbitrary pswitch set $S=\{s_1,s_2,\dots,s_{|S|}\}$. In this case,
  we propose an algorithm based on local optimization to approximate any desired probability. The analysis reveals that
  the approximation error of a switching circuit decreases exponentially with an increasing circuit size.
\end{abstract}


\IEEEpeerreviewmaketitle

\section{Introduction}

\IEEEPARstart{I}{n} his master's thesis of 1938, Claude Shannon
demonstrated how Boolean algebra can be used to synthesize and
simplify relay circuits, establishing the foundation of modern digital
circuit design \cite{Shannon1938}.
Later, deterministic switches were
replaced with probabilistic switches to make stochastic switching
circuits, which were studied in \cite{Wilhelm2008}. There are a few features of stochastic switching circuits
that make them very similar to neural systems. First, randomness is inherent in neural systems and it may play a crucial role in thinking and reasoning. Switching (and relaying) technique provides us a natural way of manipulating this randomness. Second, in a switching system, each switch can be treated as either a memory element or a control element for computing. This might enable creating an intelligent system where storage and computing are highly integrated. In this paper, we study stochastic switching circuits from a basic starting point with focusing on probability synthesis. We
consider two-terminal stochastic switching circuits, where each
probabilistic switch, or \emph{pswitch}, is closed with some
probability chosen from a finite set of rational numbers, called a
\emph{pswitch set}. By selecting pswitches with different probabilities and composing
them in appropriate ways, we can realize a variety of different closure probabilities.

\begin{figure}[!t]
\centerline{\subfigure[ssp circuit. $P(C)=\frac{5}{8}$.]{\includegraphics[width=1.5in]{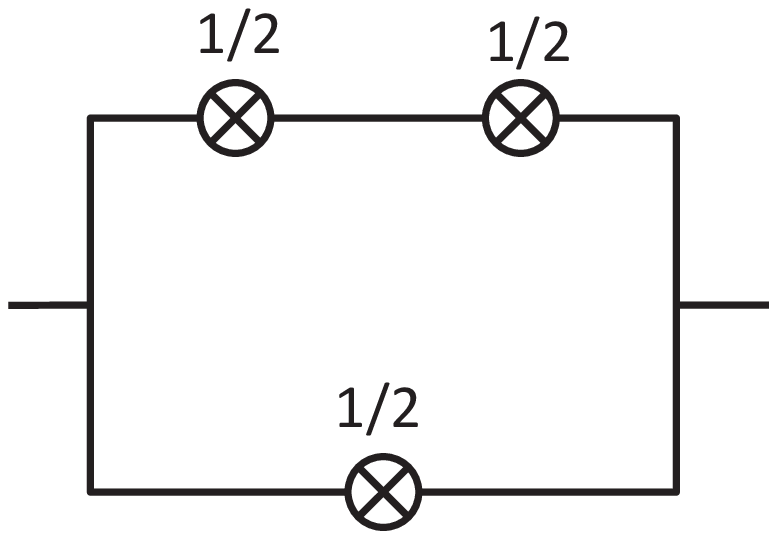} }
\hspace{0.5cm}
 \subfigure[sp circuit,
non-ssp. \newline $P(C)=\frac{7}{16}$.]{\includegraphics[width=1.5in]{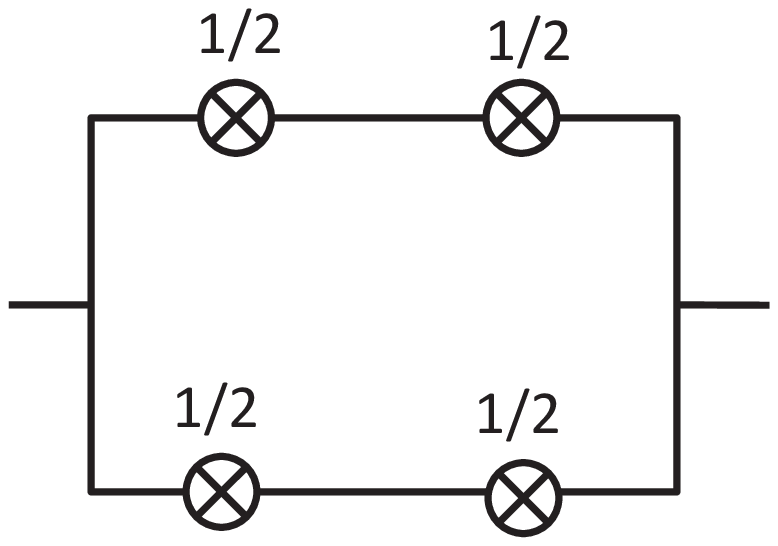}}}
\centerline{
 \subfigure[
non-sp circuit. $P(C)=\frac{1}{2}$.]{\includegraphics[width=1.5in]{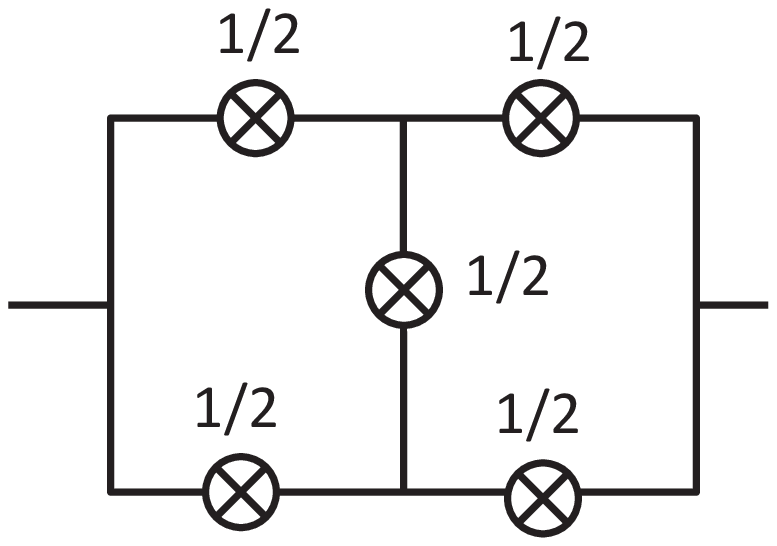} }}
\caption{Examples of ssp, sp, and non-sp
  circuits.} \label{fig_circuitexamples}
\end{figure}

Formally, for a two-terminal stochastic switching circuit $C$, the probabilities of pswitches are taken from a fixed pswitch set $S$, and all these pswitches are open or closed independently. We use $P(C)$ to denote the probability that the two terminals of $C$ are
connected, and call $P(C)$ the \emph{closure probability} of $C$. Given a pswitch set $S$, a probability $x$ can be \emph{realized} if and only if there exists a
circuit $C$ such that $x=P(C)$. Based on the ways of composing pswitches, we have
series-parallel (sp) circuits and non-series-parallel (non-sp) circuits.
An sp circuit consists of either a single pswitch or two sp circuits
connected in series or parallel, see the circuit in Fig.~\ref{fig_circuitexamples}(a) and \ref{fig_circuitexamples}(b) as examples. The circuit in Fig.~\ref{fig_circuitexamples}(c) is a non-sp circuit. A special type of sp circuits is called simple-series-parallel (ssp) circuits.  An ssp circuit is either a
single pswitch, or is built by taking an ssp circuit and adding another pswitch in either series or parallel. For example, the circuit in Fig.~\ref{fig_circuitexamples}(a) is an ssp circuit but the one in Fig.~\ref{fig_circuitexamples}(b) is not.

In this paper, we first study the robustness of different stochastic switching circuits
in the presence of small error perturbations. We assume that
the probabilities of individual pswitches are taken from a fixed
pswitch set with a given error allowance of $\epsilon$; that is, the error
probabilities of the pswitches are bounded by $\epsilon$. We show that
ssp circuits are robust to small error perturbations, but the error
probability of a general sp circuit may be amplified by adding
additional pswitches. These results might help us understand why
local errors do not accumulate in a natural system, and how to enhance
the robustness of a system when designing a circuit.

Next, we study the problem of synthesizing desired probabilities with stochastic switching circuits.
We mainly focus on ssp circuits due to their robustness
against small error perturbations. Two main questions are addressed:
(1) \emph{Expressibility}: Given the pswitch set $S=\{\frac{1}{q},\frac{2}{q},\ldots,\frac{q-1}{q}\}$, where $q$ is an
  integer, what kind of probabilities can be realized using stochastic switching circuits?
  And how many pswitches are sufficient to realize them?
(2) \emph{Approximation}:  Given an arbitrary pswitch set $S$, how can we construct
a stochastic switching circuit using as a few as possible pswitches, to get a good approximation of
the desired probabilities?

The study of probability synthesis based on stochastic switching circuits has widespread applications. Recently,
people found that DNA molecules can be constructed that closely approximate
the dynamic behavior of arbitrary systems of coupled chemical reactions \cite{Soloveichik10}, which leads to the field of molecular computing \cite{Cook09}.
In such systems, the quantities of
molecules involved in a reaction are often surprisingly
small, and the exact sequence of reactions is determined by chance
\cite{Fett2007}. Stochastic switching circuits provide a simple and powerful tool to manipulate stochasticity in molecular systems.
Comparing with combinational logic circuits, stochastic switching circuits are easier to implement using molecular reactions. Another type of applications is probabilistic electrical systems without sophisticated computing components. In such systems, stochastic switching circuits have many advantages in generating desired probabilities,
including its constructive simplicity, robustness, and low cost.

The remainder of this paper is organized as follows: Section
\ref{section_relatedworks} describes related work and introduces some existing
results on stochastic switching circuits.  In Section \ref{section_robustness},
we analyze the robustness of different kinds of stochastic switching
circuits. Then we discuss the expressibility of
stochastic switching circuits in Section \ref{section_scenario1} and probability
approximation in Section \ref{section_scenario2}, followed by the conclusion in Section \ref{switch_section_conclusion}.

\section{Related Works and Preliminaries}
\label{section_relatedworks}

There are a number of studies related to the problem of generating desired distributions from the algorithmic perspective. This problem dates back to von Neumann \cite{Neumann1951}, who considered of simulating an unbiased coin using a biased coin with unknown probability. Later, Elias \cite{Elias1972} improved this algorithm such that the expected number of unbiased random bits generated per coin toss is asymptotically equal to the entropy of the biased coin. On the other hand, people have considered the case that the probability distribution of the tossed coin is known. Knuth and Yao
\cite{Knuth1976} have given a procedure to generate an arbitrary
probability using an unbiased coin. Han and Hoshi \cite{Han1997} have demonstrated how to
generate an arbitrary probability using a general $M$-sided biased coin. All these works aim to efficiently convert one distribution to another. However, they require computing models and may not be applicable for
some simple or distributed electrical/molecular systems.

There are a number of studies focusing on synthesizing a simple physical device
to generate desired probabilities. Gill \cite{Gill62}\cite{Gill63} discussed
the problem of generating rational probabilities using a sequential
state machine. Motivated by neural computation, Jeavons et al. provided an algorithm
to generate binary sequences with probability $\frac{a}{q^n}$ from a set of stochastic binary sequences
with probabilities in $\{\frac{1}{q},\frac{2}{q},\ldots,\frac{q-1}{q}\}$ \cite{Jeavons94}. Their method
can be implemented using the concept of linear feedback shift registers. Recently, inspired by PCMOS technology \cite{Chakrapani2007}, Qian et al. considered the synthesis of decimal probabilities
using combinational logic \cite{Qian2011}. They have considered three different scenarios,
depending on whether the given probabilities can be duplicated, and
whether there is freedom to choose the probabilities. In contact to the foregoing contributions, we consider
the properties and probability synthesis of stochastic switching circuits.
Our approach is orthogonal and complementary to that of Qian and Riedel, which is based on combinational logic. Generally,
each switching circuit can be equivalently expressed by a combinational logic circuit.
All the constructive methods of stochastic switching circuits in this paper can be directly applied to probabilistic combinational logic circuits.

In the rest of this section, we introduce the original work that started the study on stochastic switching circuits (Wilhelm and Bruck \cite{Wilhelm2008}). Similar to resistor circuits \cite{MacMahon1892}, connecting one
terminal of a switching circuit $C_1$ (where $P(C_1) = p_1$) to one
terminal of a circuit $C_2$ (where $P(C_2) = p_2$) places them in
series. The resulting circuit is closed if and only if both of $C_1$
and $C_2$ are closed, so the probability of the resulting circuit is
$$p_{\mathrm{series}}=p_1\cdot p_2.$$
Connecting both terminals of $C_1$ and $C_2$ together places the
circuits in parallel. The resulting circuit is closed if and only if
either $C_1$ or $C_2$ is closed, so the probability of the resulting
circuit is $$p_{\mathrm{parallel}}=1-(1-p_1)(1-p_2)=p_1+p_2-p_1p_2.$$ Based on
these rules, we can calculate the probability of any given ssp or sp
circuit.
For example, the probability of the circuit in
Fig.~\ref{fig_circuitexamples}(a) is
$$p_{(a)}=\left(\frac{1}{2} \cdot \frac{1}{2}\right)+\frac{1}{2}-\left(\frac{1}{2}\cdot\frac{1}{2}\right)\frac{1}{2}=\frac{5}{8},$$
and the probability of the circuit in
Fig.~\ref{fig_circuitexamples}(b) is
$$p_{(b)}=\left(\frac{1}{2}\cdot\frac{1}{2}\right)+\left(\frac{1}{2}\cdot\frac{1}{2}\right)-\left(\frac{1}{2}\cdot\frac{1}{2}\right)\left(\frac{1}{2}\cdot\frac{1}{2}\right)=\frac{7}{16}.$$

Let us consider the non-sp circuit in Fig.~\ref{fig_circuitexamples}(c). In this circuit, we call the pswitch
in the middle a `bridge'. If the bridge is closed, the circuit has a closure probability of $\frac{9}{16}$. If the bridge is open, the
circuit has a closure probability of
$\frac{7}{16}$. Since the bridge is closed with probability
$\frac{1}{2}$, the overall probability of the circuit is
$$p_{(c)}=\frac{1}{2}\cdot\frac{9}{16}+\frac{1}{2}\cdot\frac{7}{16}=\frac{1}{2}.$$

An important and interesting question is that if $S$ is uniform, i.e., $S=\{\frac{1}{q}, \frac{2}{q}, \ldots,\frac{q-1}{q}\}$ for some $q$, what kind of probabilities can be realized using stochastic switching circuits?
In \cite{Wilhelm2008}, Wilhelm and Bruck proposed an optimal algorithm (called B-Algorithm) to realize
all rational probabilities of the form $\frac{a}{2^n}$ with
$0<a<2^n$, using an ssp circuit when $S = \{\frac{1}{2}\}$. In their algorithm, at most $n$ pswitches are used, which is optimal.
They also proved that given the
pswitch set $S=\{\frac{1}{3},\frac{2}{3}\}$, all rational
probabilities $\frac{a}{3^n}$ with $0<a<3^n$ can be realized by an
ssp circuit with at most $n$ pswitches; given the pswitch set
$S=\{\frac{1}{4},\frac{2}{4},\frac{3}{4}\}$, all rational
probabilities $\frac{a}{4^n}$ with $0<a<4^n$ can be realized by an
ssp circuit with at most $2n-1$ pswitches.

\begin{figure}[!ht]
  \centerline{\subfigure[Initial
    circuit. $P=\frac{1}{4}$.]{\includegraphics[width=1.6in]{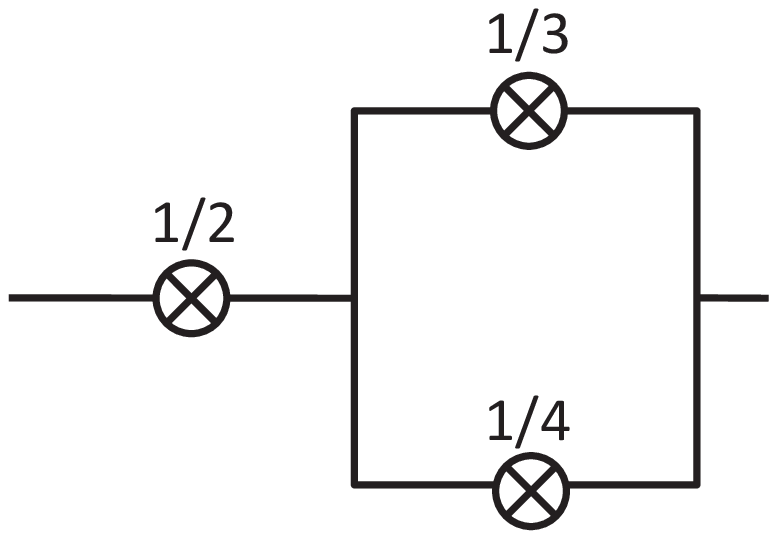}
    } \subfigure[The
    dual. $P=\frac{3}{4}$.]{\includegraphics[width=1.6in]{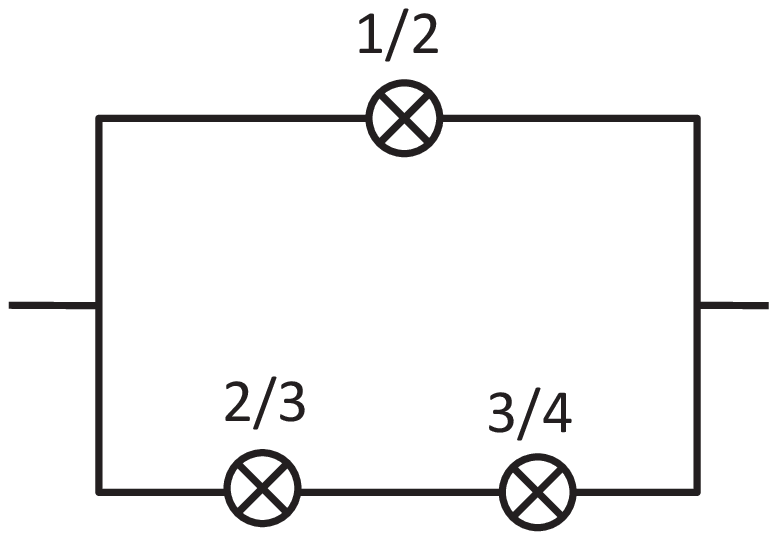}
    }}
\caption{A circuit and its dual.} \label{fig_duality}
\end{figure}

Wilhelm and Bruck also demonstrated the concept of \emph{duality} in sp
circuits. The dual of a single pswitch of probability $p$ appearing in
series is the corresponding pswitch of probability $1-p$ appearing in
parallel. Similarly, the dual of a pswitch of probability $p$
appearing in parallel is a pswitch of probability $1-p$ appearing in
series. For example, in Fig.~\ref{fig_duality}, the circuit in (b) is the dual of
the circuit in (a), and vice versa. It can be proved that dual circuits satisfy the
following relation:

\begin{Theorem}[Duality Theorem \cite{Wilhelm2008}]
  For a stochastic series-parallel circuit $C$ and its dual
  $\overline{C}$, we have
$$P(C)+P(\overline{C})=1,$$
where $P(C)$ is the probability of circuit $C$ and $P(\overline{C})$
is the probability of circuit $\overline{C}$.
\end{Theorem}

\section{Robustness}
\label{section_robustness}

In this section, we analyze the robustness of different kinds of
stochastic switching circuits, where the probabilities of individual
pswitches are taken from a fixed pswitch set, but given an error
allowance of $\epsilon$; i.e., the error probabilities of the
pswitches are bounded by $\epsilon$. For a stochastic circuit with
multiple pswitches, the \emph{error probability} of the circuit is the
absolute difference between the probability that the circuit is closed
when error probabilities of pswitches are included, and the
probability that the circuit is closed when error probabilities are
omitted. We show that ssp circuits are robust to small error
perturbations, but the error probability of a general sp circuit may
be amplified with additional pswitches.

\subsection{Robustness of ssp Circuits}

Here, we analyze the susceptibility of ssp circuits to small
error perturbations in individual pswitches. Based on our assumption, instead of assigning a
pswitch a probability of $p$, the pswitch may be assigned a
probability between $p-\epsilon$ and $p+\epsilon$, where $\epsilon$ is
a fixed error allowance.

\begin{Theorem}[Robustness of ssp circuits]
  Given a pswitch set S, if the error probability of each pswitch is
  bounded by $\epsilon$, then the total error probability of an ssp
  circuit is bounded by
$$\frac{\epsilon}{\min(\min(S),1-\max(S))}.$$
\end{Theorem}

\proof We induct on the number of pswitches. If we have just one
pswitch, the result is trivial. Suppose the result holds for $n$
pswitches, and note that for an ssp circuit with $n+1$ pswitches, the
last pswitch will either be added in series or in parallel with the
first $n$ pswitches. By the induction hypothesis, the circuit
constructed from the first $n$ pswitches has probability
$p+\epsilon_1$ of being closed, where $\epsilon_1$ is the error
probability introduced by the first $n$ pswitches and
$|\epsilon_1|\leq \frac{ \epsilon }{\min(\min(S),1-\max(S))}$.
The $(n+1)$st pswitch has probability $t+\epsilon_2$ of being closed,
where $t\in S$ and $|\epsilon_2| \leq \epsilon$.

\begin{figure}[!ht]
  \centerline{\subfigure[The last pswitch is added in
    series.]{\includegraphics[width=1.5in]{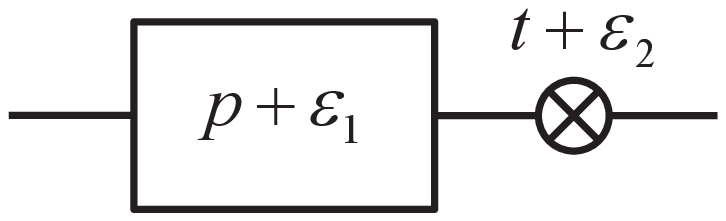}}
    \hspace{0.3in}
    \subfigure[The last pswitch is added in
    parallel.]{\includegraphics[width=1.5in]{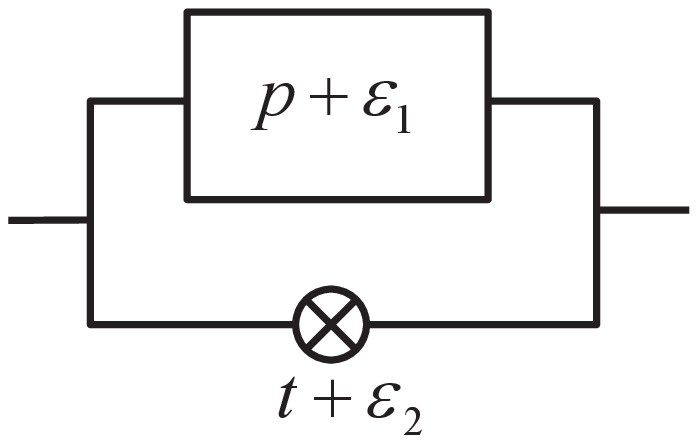}
    }}
\caption{Robustness of ssp circuits.} \label{fig_robustssp}
\end{figure}

If the $(n+1)$st pswitch is added in series, see Fig.~\ref{fig_robustssp}(a), then the new circuit
(with errors) has probability
$$(p+\epsilon_1)(t+\epsilon_2)=tp+\epsilon_2(p+\epsilon_1)+t\epsilon_1$$
of being closed. Without considering the error probability of each
pswitch, the probability of the new circuit is $tp$.
Hence, the overall error probability of the circuit is
$e_1=\epsilon_2(p+\epsilon_1)+t\epsilon_1$.
By the triangle inequality and the induction hypothesis,
\begin{eqnarray*}
 |e_1|
   &\leq& |\epsilon_2||(p+\epsilon_1)|+t|\epsilon_1|
  \leq  |\epsilon_2|+t|\epsilon_1|\\
  &\leq & \left(\frac{t}{\min(\min(S),1-\max(S))}+1\right)\epsilon\\
  &\leq & \frac{\min(\min(S),1-\max(S))+\max(S)}{\min(\min(S),1-\max(S))}\cdot\epsilon\\
  &\leq & \frac{\epsilon}{\min(\min(S),1-\max(S))},
\end{eqnarray*}
completing the induction.

Similarly, if the $(n+1)$st pswitch is added in parallel, see Fig.~\ref{fig_robustssp}(b), then the
new circuit (with errors) has probability
\begin{eqnarray*}
  && (p+\epsilon_1)+(t+\epsilon_2)-(p+\epsilon_1)(t+\epsilon_2)\\
  &=& (p+t-tp)+\epsilon_1(1-t)+\epsilon_2(1-p-\epsilon_1)
\end{eqnarray*}
of being closed. Without considering the error probability of each
pswitch, the probability that the circuit is closed is $p+t-tp$.
Hence, the overall error probability of the circuit with $n+1$
pswitches is $e_2=\epsilon_1(1-t)+\epsilon_2(1-p-\epsilon_1)$. Again
using the induction hypothesis and the triangle inequality, we have
\begin{eqnarray*}
  &&|e_2| \\
  &\leq& |\epsilon_2||(1-p-\epsilon_1)|+(1-t)|\epsilon_1| \\
  &\leq&  |\epsilon_2|+(1-t)|\epsilon_1|\\
  &\leq & \left(\frac{1-t}{\min\{\min(S),1-\max(S))}+1\right)\epsilon\\
  &\leq & \frac{\min\{\min(S),1-\max(S))+1-\min(S)}{\min\{\min(S),1-\max(S))}\cdot\epsilon\\
  &\leq & \frac{\epsilon}{\min\{\min(S),1-\max(S))}.
\end{eqnarray*}
This completes the proof.
\hfill\QED

The theorem above implies that ssp circuits are robust to small error perturbations:
no matter how big the circuit is, the error probability of an ssp circuit
will be well bounded by a constant times $\epsilon$.
Let us consider a case that $S = \{\frac{1}{ 2}\}$. In this case,
the overall error probability of any ssp circuit is bounded by
$2\epsilon$ if each pswitch is given an error allowance of $\epsilon$.

\subsection{Robustness of sp Circuits}

We have proved that for a given pswitch set $S$, the overall error
probability of an ssp circuit is well bounded. We want to know whether this property holds for all sp
circuits. Unfortunately, we show that as the number of pswitches increases,
the overall error probability of an sp circuit may also increase. In this subsection, we will
give the upper bound and lower bound for the error probabilities of sp circuits.

\begin{Theorem}[Lower bound for sp circuits]
  Given a pswitch set $S$, if the error probability of each pswitch is
  $\epsilon$ (where $\epsilon\rightarrow0$), then there exists an sp
  circuit of size $n$ with overall error probability $O(\log
  n)\epsilon$.
\end{Theorem}

\begin{figure}[!ht]
\centering
\includegraphics[width=3.5in]{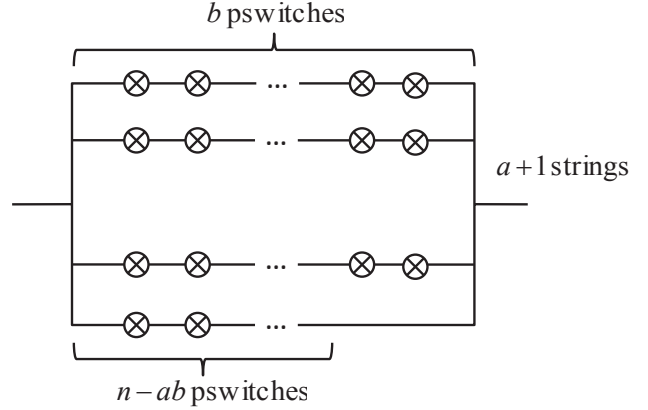}
\caption{The construction of an sp circuit.}
\label{fig_spbound}
\end{figure}

\proof  Suppose $p\in S$, and without loss of generality, assume
$\epsilon>0$. We construct an sp circuit as shown in
Fig.~\ref{fig_spbound}, by connecting $a+1$ strings of pswitches in
parallel. Among these strings, we have $a$ strings of $b$ pswitches
and one string of $n-ab$ pswitches, and all pswitches have probability
$p$. Now, we let $a$ and $b$ satisfy the following
relation: $$a=\left\lceil\frac{n}{b}\right\rceil-1,
a=\left\lfloor(\frac{1}{p})^b\right\rfloor.$$
Without considering pswitch errors, the probability of the circuit is
$$p_1=1-(1-p^b)^a(1-p^{n-ab}).$$

Suppose we introduce an error of $\epsilon$ to each pswitch, such that
the probability of each pswitch is $p+\epsilon$ (assume $\epsilon>0$). Then the probability of the circuit is
$$p_2(\epsilon)=1-(1-(p+\epsilon)^b)^a(1-(p+\epsilon)^{n-ab}),$$
where $p_2(0)=p_1$.

Assuming $n$ is large enough, we have the following error probability
for the circuit:
\begin{eqnarray*}
  &&e_1 \\
  &=& p_2(\epsilon)-p_1 \\
  &\simeq & p_2'(\epsilon)\epsilon\\
  &\simeq& -[(1-(p+\epsilon)^b)^a(1-(p+\epsilon)^{n-ab})]'\epsilon\\
  &\simeq& -[e^{-a(p+\epsilon)^b}(1-(p+\epsilon)^{n-ab})]'\epsilon\\
  &\simeq& e^{-a(p+\epsilon)^b} ab(p+\epsilon)^{b-1}(1-(p+\epsilon)^{n-ab})\epsilon\\
  & & + e^{-a(p+\epsilon)^b} (n-ab) (p+\epsilon)^{n-ab-1}\epsilon\\
  &\simeq& [e^{-ap^b} ab p^{b-1} (1- p^{n-ab})+e^{-ap^b}(n-ab)p^{n-ab-1}]\epsilon\\
  &\simeq& [e^{-1}\frac{b}{p}(1-p^{n-ab})+ e^{-1}(n-ab)p^{n-ab-1}]\epsilon.
\end{eqnarray*}

So when $n$ is large enough, we have
$$ e^{-1}\frac{1-p}{p}b\epsilon \leq |e_1|\leq e^{-1}\frac{1}{p}b\epsilon.$$

Since $b\lfloor(\frac{1}{p})^b\rfloor<n\leq
b(\lfloor(\frac{1}{p})^b\rfloor+1)$ for large $n$, we have
$$b\sim \frac{\log n}{\log \frac{1}{p}}-\frac{\log\log n}{\log \frac{1}{p}} +\frac{\log\log\frac{1}{p}}{\log\frac{1}{p}}\sim \frac{\log n}{\log \frac{1}{p}}.$$

Finally, we have $|e_1|\sim O(\log n)\epsilon$, completing the proof.
\hfill\QED

In the following theorem, we will give the upper bound for the error probabilities of
sp circuits.

\begin{Theorem}[Upper bound for sp circuits]
  Given an sp circuit with $n$ pswitches taken from a finite pswitch
  set $S$, if each pswitch has error probability  bounded by
  $\epsilon$, then the total error probability of the circuit is
  bounded by $ c\sqrt{n}\epsilon$,
  where $c=\max_{t\in S}\frac{1}{\sqrt{t(1-t)}}$ is a constant.\label{robustness_sp}
\end{Theorem}

\proof
Assume $x$ is a pswitch in a stochastic circuit $C$, and the actual probability of
$x$ is $t_x+\epsilon_x$, where $\epsilon_x$ is the error part such that
$|\epsilon_x|\leq \epsilon$.  Let $P(C|x=1)$ denote
the probability of circuit $C$ when $x$ is closed, and let $P(C|x=0)$
denote the probability of $C$ when $x$ is open.

Without considering the error probability of $x$, the probability of
circuit $C$ can be written as
$$P_{x}(C)=t_x P(C|x=1) +(1-t_x)P(C|x=0).$$
Considering the error part of $x$, we have
$$P(C)=(t_x+\epsilon_x) P(C|x=1) + (1-t_x-\epsilon_x)P(C|x=0).$$

In order to prove the theorem, we define a term called the
\emph{error contribution}.  In a circuit $C$, the error contribution of
pswitch $x$ is defined as
$$e_x(C)=|P(C)-P_x(C)|=\epsilon_x|P(C|x=1)-P(C|x=0)|$$
$$\leq \epsilon(P(C|x=1)-P(C|x=0)).$$

In the rest of the proof, we have two steps.

(1) In the first step, we show that given an sp circuit with size $n$,
there exists at least one pswitch such that its error contribution is
bounded by $\frac{c\sqrt{(1-P)P}}{\sqrt{n}}\epsilon$,
where $P$ is the probability of the sp circuit and $c=\max_{t\in S}
  \frac{1}{\sqrt{t(1-t)}}$.

We induct on the number of pswitches. If the circuit has only
one pswitch, the result is trivial. Suppose the result holds for $k$
pswitches for all $k<n$. We need to prove that the result holds for any sp
circuit $C$ with $n$ pswitches.

Suppose circuit $C$ is constructed by connecting two sp circuits
$C_1$ and $C_2$ in series, where $C_1$ has $n_1$ pswitches and
probability $P_1$, and $C_2$ has $n_2$ pswitches and probability
$P_2$. Note that $n_1+n_2=n$ and $n_1<n, n_2<n$.

By the induction hypothesis, circuit $C_1$ contains a pswitch $x_1$
with error contribution
$$e_{x_1}(C_1)\leq \frac{c\sqrt{(1-P_1)P_1}}{\sqrt{n_1}}\epsilon.$$

In circuit $C$, the error contribution of pswitch $x_1$ is
$$e_{x_1}(C)=|P(C)-P_{x_1}(C)|=P_2|P(C_1)-P_{x_1}(C_1)|$$
$$=P_2 e_{x_1}(C_1).$$

Similarly, in the circuit $C_2$, there exists a pswitch
$x_2$ such that the error contribution of $x_2$ is
$$e_{x_2}(C_2)\leq \frac{c\sqrt{(1-P_2)P_2}}{\sqrt{n_2}}\epsilon,$$
and the error contribution of $x_2$ to circuit $C$ is
$$e_{x_2}(C)=P_1 {e}_{x_2}(C_2).$$

Since the circuit $C$ is constructed by connecting circuits $C_1$ and
$C_2$ in series, the probability of circuit $C$ is $P=P_1P_2$. Thus,
we only need to prove that either $e_{x_1}(C)$ or $e_{x_2}(C)$ is
bounded by
$$\frac{c\sqrt{(1-P_1P_2)P_1P_2}}{\sqrt{n_1+n_2}}\epsilon,$$
This can be proved by contradiction as follows.

Assume both $e_{x_1}(C)$ and $e_{x_2}(C)$ are larger than
$\frac{c\sqrt{(1-P_1P_2)P_1P_2}}{\sqrt{n_1+n_2}}\epsilon$. Then we have
\[
P_2 \frac{c\sqrt{(1-P_1)P_1}}{\sqrt{n_1}} >
\frac{c\sqrt{(1-P_1P_2)P_1P_2}}{\sqrt{n_1+n_2}}
\]
and
\[
P_1 \frac{c\sqrt{(1-P_2)P_2}}{\sqrt{n_2}} >
\frac{c\sqrt{(1-P_1P_2)P_1P_2}}{\sqrt{n_1+n_2}},
\]
which can be simplified as
\[
\frac{n_1}{n_1+n_2} < \frac{(1-P_1)P_2}{1-P_1P_2}
\]
and
\[
\frac{n_2}{n_1+n_2} < \frac{(1-P_2)P_1}{1-P_1P_2}.
\]

Adding the two inequalities yields
$$P_1+P_2-1-P_1P_2=-(1-P_1)(1-P_2)>0,$$
which is a contradiction.
So we conclude that at least one of $e_{x_1}(C)$ and
$e_{x_2}(C)$ is bounded by
$\frac{c\sqrt{(1-P_1P_2)P_1P_2}}{\sqrt{n_1+n_2}}\epsilon$ when
$C$ is constructed by connecting two sp circuits in series.
If the circuit $C$ is constructed by connecting two sp circuits in parallel,
using a similar argument, we can get the same conclusion.

Finally,  we get that given an sp circuit with size $n$,
there exists at least one pswitch such that its error contribution is
bounded by $\frac{c\sqrt{(1-P)P}}{\sqrt{n}}\epsilon$.

(2) In the second step, we prove the theorem based on the result above.

We again induct on the number of pswitches. If we have less than three
pswitches, the result is trivial. Suppose the result holds for any sp circuit with $n\geq
2$ pswitches; we want to prove that the result also holds for any circuit with $n+1$
pswitches.

Based on the result in the first step, we know that given an sp circuit $C$
with $n+1$ pswitches, there exists a pswitch $x$ with error contribution
bounded by $\frac{c}{2\sqrt{n+1}}\epsilon$.

\begin{figure}[!ht]
  \centerline{\subfigure[Circuit $C$.]{\includegraphics[width=1.4in]{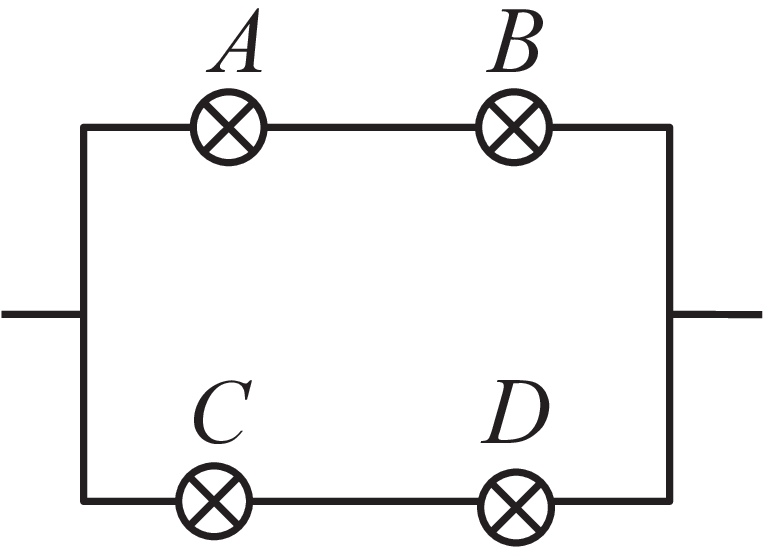}} \hspace{0.1in}
    \subfigure[$D_1$, $A$ closed.]{\includegraphics[width=1.4in]{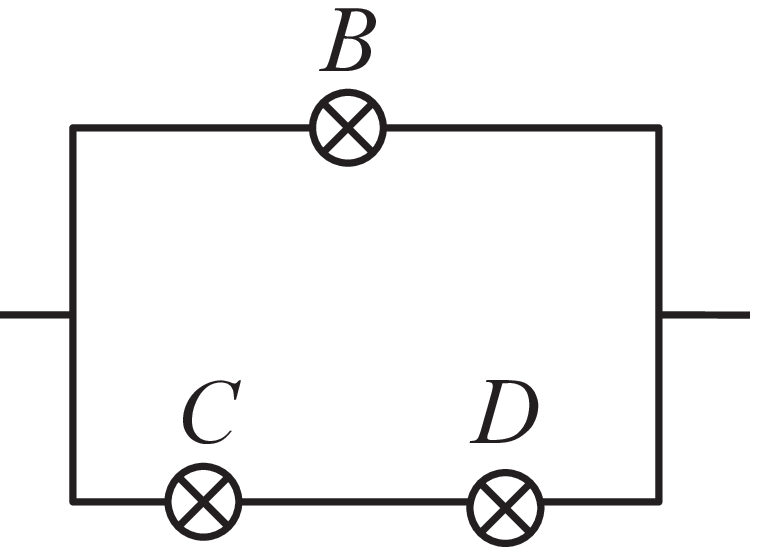}}}
    \centerline{
    \subfigure[$D_2$, $A$  open.]{\includegraphics[width=1.4in]{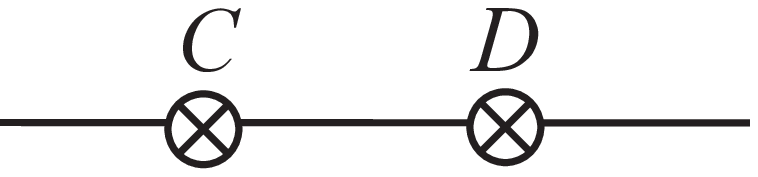}}}
\caption{An illustration of keeping a pswitch $A$ closed or open in an sp circuit $C$.} \label{fig_robustsp_2}
\end{figure}

By keeping pswitch $x$ closed, we obtain an sp circuit $D_1$ with at
most $n$ pswitches. Please see Fig.~\ref{fig_robustsp_2}(a)(b) as an example.
Without considering pswitch errors, $D_1$ is closed with probability $p_1$;
considering all pswitch errors, $D_1$ is closed with probability
$q_1$. According to our assumption, we have
$$e_1=|q_1-p_1|\leq c\sqrt{n}\epsilon.$$

By keeping pswitch $x$ open, we obtain an sp circuit $D_2$ with at
most $n$ pswitches.  Please see Fig.~\ref{fig_robustsp_2}(a)(c) as an example.  Without considering pswitch errors, $D_2$ is
closed with probability $p_2$; considering all pswitch errors,
$D_2$ is closed with probability $q_2$. According to our assumption,
we have
$$e_2=|q_2-p_2|\leq c\sqrt{n}\epsilon.$$

For the initial sp circuit $C$ with $n+1$ pswitches, without considering
pswitch errors, the overall probability of the circuit is given
by $$t_xp_1+(1-t_x)p_2,$$ where $t_x$ is the probability of pswitch $x$.

Considering all pswitch errors, the overall probability of the circuit is
$$(t_x+\epsilon_x)q_1+(1-t_x-\epsilon_x) q_2.$$

We know that the error contribution of pswitch
$x$ to the circuit $C$ is
$$e_x(C)=\epsilon_x|q_2-q_1| \leq \frac{c}{2\sqrt{n+1}}\epsilon.$$

Then by the triangle inequality, we can get the error probability of the circuit $C$:
\begin{eqnarray*}
  &&e \\
  &=& |(t_x+\epsilon_x)q_1+(1-t_x-\epsilon_x) q_2-(t_xp_1+(1-t_x)p_2)|\\
  &\leq & t_x|q_1-p_1| + (1-t_x) |q_2-p_2| +\epsilon_x|q_2-q_1|\\
  &\leq &  \frac{c\sqrt{n(n+1)}+ \frac{c}{2}}{\sqrt{n+1}}\epsilon\\
  &\leq & c\frac{(n+\frac{1}{2})+\frac{1}{2}}{\sqrt{n+1}}\epsilon\\
  &=& c\sqrt{n+1}\epsilon.
\end{eqnarray*}
This finishes the induction.
\hfill\QED

\subsection{Robustness of Non-sp Circuits}

Here we extend our discussion to the case of general
stochastic switching circuits. We have the following theorem, which
clearly holds for sp and ssp circuits:

\begin{Theorem}[Upper bound for general circuits]
  Given a general stochastic switching circuit with $n$ pswitches
  taken from a finite pswitch set $S$, if each pswitch has error
  probability bounded by $\epsilon$, then the total probability of the
  circuit is bounded by $n\epsilon$.
\end{Theorem}

\begin{figure}[!ht]
\centering
\includegraphics[width=1.8in]{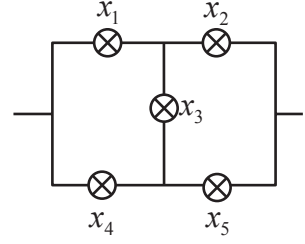}
\caption{An example of a general stochastic switching circuit.}
\label{fig_generalbound}
\end{figure}

\proof We first index all the pswitches in the circuit $C$ as $x_1, x_2, \ldots, x_n$, see
Fig.~\ref{fig_generalbound} as an example.

Let $t_i+\epsilon_i$ be the probability that $x_i$ is closed, where $\epsilon_i$
is the error part such that $|\epsilon_i|\leq \epsilon$.
Let $P^{(k)}$ denote the probability that $C$ is closed when we only
take into account the errors of $x_1, x_2, \ldots,x_k$, i.e.,
$$P^{(k)}=P(t_1+\epsilon_1,\dots,t_k+\epsilon_k, t_{k+1},\dots,t_n),$$
where $P(a_1,a_2, \ldots,a_n)$ indicates the probability of $C$ if $x_i$ is closed with probability $a_i$
for all $1\leq i\leq n$.

The overall error probability of the circuit $C$ can then be written as
\begin{eqnarray*}
  e &=& P^{(n)}-P(0)\\
  &=&  (P^{(n)}-P^{(n-1)})+(P^{(n-1)}-P^{(n-2)})+ \cdots\\
  && + (P^{(1)}-P^{(0)}).
\end{eqnarray*}

Now, we prove that $|P^{(k)}-P^{(k-1)}|\leq \epsilon$ for all
$1\leq k\leq n$
\begin{eqnarray*}
&& |P^{(k)} - P^{(k-1)}| \\
  &=& |P(t_1+\epsilon_1,\dots,t_k+\epsilon_k, t_{k+1},\dots,t_n)\\
  && - P(t_1+\epsilon_1,\dots,t_{k-1}+\epsilon_{k-1}, t_{k},\dots,t_n)|\\
  &=& |(t_k+\epsilon_k)P(t_1+\epsilon_1,\dots,1, t_{k+1},\dots,t_n)\\
  && +(1-t_k-\epsilon_k)P(t_1+\epsilon_1,\dots,0, t_{k+1},\dots,t_n)\\
  &&-t_k P(t_1+\epsilon_1,\dots,t_{k-1}+\epsilon_{k-1}, 1,\dots,t_n)\\
  && - (1-t_k) P(t_1+\epsilon_1,\dots,t_{k-1}+\epsilon_{k-1}, 0,\dots,t_n)|\\
  &=& |\epsilon_k[P(t_1+\epsilon_1,\dots,1, t_{k+1},\dots,t_n)\\
  &&-P(t_1+\epsilon_1,\dots,0, t_{k+1},\dots,t_n)]|\\
  &\leq & \epsilon.
\end{eqnarray*}

Therefore, we have
\[
e \leq \sum_{k=1}^{n}|P^{(k)}-P^{(k-1)}|\leq n\epsilon,
\]
as we wanted. \hfill\QED

Note that in most of cases, the actual error probability of a circuit is much smaller than $n\epsilon$ when $n$ is large.
However, $n\epsilon$ is still achievable in the following case: by placing $n$ pswitches with probability $p-\epsilon$ in series, where $\epsilon\rightarrow \infty$, we can get a circuit
whose probability is
$$(p-\epsilon)^n \approx p^n - np^{n-1}\epsilon.$$
Without considering the errors, the probability of the
circuit is $p^n$, so the overall error is $n\cdot p^{n-1}\epsilon$.
Choosing $p$ sufficiently close to $1$, we can make the error
probability of the circuit arbitrarily close to $n\epsilon$.

\section{Expressibility}
\label{section_scenario1}

\begin{table*}[!t]
\renewcommand{\arraystretch}{1.3}
\centering{
\caption{The expressibility of stochastic switching circuits}
\begin{tabular}{|c|c|c|c|}
  \hline
  $S=\{\frac{1}{q},\frac{2}{q},\dots,\frac{q-1}{q}\}$  & Can all $\frac{a}{q^n}$ be realized? & upper bound of circuit size\\
  \hline & &  \\[-1.3em]\hline
  $q$ is even & yes, ssp circuit & $\lceil\log_2 q\rceil(n-1)+1$\\
  \hline
  $q$ is an odd multiple of 3 & yes, ssp circuit & $\lceil\log_3 q\rceil(n-1)+1$\\
  \hline
  $q$ is a prime number larger than $3$  & no, not by sp circuits & --\\
  \hline
  other values of $q$  & open problem & --\\
  \hline
\end{tabular}
\label{table_expressibility1}}
\end{table*}

In the previous section, we showed that ssp circuits are robust against noise.
This property is important in natural systems and useful in engineering system design, because
the local error of a system should not be amplified. In this section, we consider another
property of stochastic switching circuits, called expressibility. Namely, given a pswitch set
$S=\{\frac{1}{q},\frac{2}{q}, \ldots,\frac{q-1}{q}\}$ for some integer $q$, the questions we ask are: What kinds of probabilities
can be realized using stochastic switching circuits (or only ssp circuits)? How many pswitches are sufficient?
Wilhelm and Bruck
\cite{Wilhelm2008} proved that if $q=2$ or $q=3$, all rational
$\frac{a}{q^n}$, with $0<a<q^n$, can be realized by an ssp circuit
with at most $n$ pswitches, which is optimal. They also showed that if
$q=4$, all rational $\frac{a}{q^n}$, with $0<a<q^n$, can be realized
using at most $2n-1$ pswitches. In this section we generalize these results:
\begin{enumerate}
  \item If $q$ is an even number, all rational $\frac{a}{q^n}$, with
  $0<a<q^n$, can be realized by an ssp circuit with at most
  $\lceil\log_2 q \rceil(n - 1)+1$ pswitches (Theorem \ref{Theorem1}).
  \item If $q$ is odd and a multiple of $3$, all rational
  $\frac{a}{q^n}$, with $0<a<q^n$, can be realized by an ssp circuit
  with at most $\lceil\log_3 q \rceil(n - 1)+1$ pswitches (Theorem \ref{Theorem2}).
  \item However, if $q$ is a prime number greater than $3$, there exists at least
one rational $\frac{a}{q^n}$, with $0<a<q^n$, that cannot be realized
using an sp circuit (Theorem \ref{theorem_switching_5}).
\end{enumerate}
Table \ref{table_expressibility1} summarizes these results. We see that when $q=2,3,$ or 4, our results agree with the results in \cite{Wilhelm2008}.

\subsection{Backward Algorithms}
\label{subsec_backwardalg}

As mentioned in \cite{Wilhelm2008}, switching circuits may be synthesized
using forward algorithms, where circuits are built by adding pswitches
sequentially, or backward algorithms, where circuits are built
starting from the ``outermost'' pswitch.


\begin{figure}[!ht]
\includegraphics[width=3.6in]{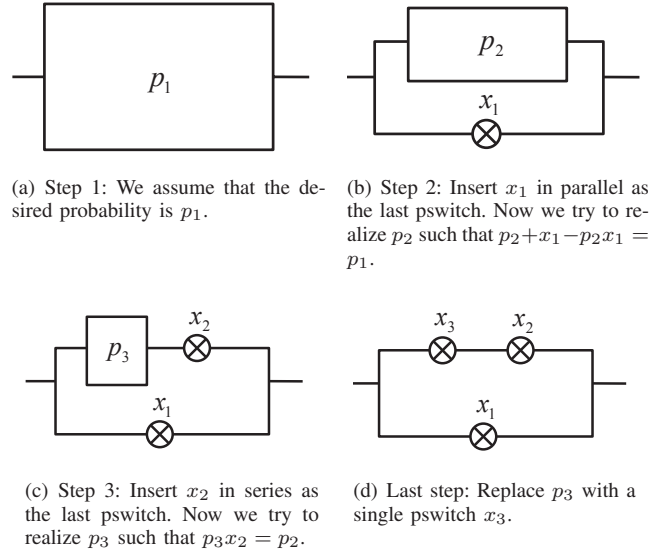}
  \caption{An example of the backward algorithm.}
\label{fig_backward}
\end{figure}

Fig.~\ref{fig_backward} gives a simple demonstration of a backward algorithm.
Assume that the desired probability is $p_1$ and we plan to insert three pswitches, namely
$x_1, x_2, x_3$ in backward direction. Here, for simplicity, we use $x_1, x_2, x_3$ to denote the closure
probabilities of the pswitches, rather than their states ($1$ or $0$).
If $x_1\leq p_1$, then $x_1$ has to be inserted
in parallel. If $x_1>p_1$, then $x_1$ has to be inserted in series. After the insertion,
we can try to realize the inner box with probability $p_2$ such that $p_2+x_1-p_2x_1=p_1$.
This process is continued recursively until for some $m$, $p_m$ can be realized with a single pswitch.
Generally, in backward algorithms, we use $x_k$ to denote the $k$th pswitch inserted in the
backward direction, and use $p_k$ to denote the probability that we
want to realize with pswitches $x_k, x_{k+1}, x_{k+2}, \ldots$

Backward algorithms have significant
advantages over forward algorithms for probability synthesis. In a forward algorithm, if we
want to add one pswitch, we have $2|S|$ choices, since each pswitch
may be added in either series or parallel. But in a backward
algorithm, if we want to insert one pswitch, we have only $|S|$
choices. That is because the insertion (series or parallel) of a pswitch $x_k$ simply depends on
the comparison of $x_k$ and $p_k$.
Therefore, backward algorithms can significantly reduce the search space, hence are more
efficient than forward algorithms. In this paper, most of the circuit constructions are
based on backward algorithms.

\subsection{Multiples of $2$ or $3$}

We consider the case that $S=\{\frac{1}{q},\frac{2}{q},\dots,\frac{q-1}{q}\}$ and $q$ is a multiple of $2$ or $3$.
We show that based on a backward algorithm, all rational $\frac{a}{q^n}$,
with $0<a<q^n$, can be realized using a bounded number of pswitches.
Before describing the details, we introduce a characteristic function called $d$ for a given probability $\frac{b}{q^w}$, that is
$$d\left(\frac{b}{q^w}\right)=\frac{q^{w-1}}{\operatorname{gcd}(b,q^{w-1})}.$$
Note that the function $d$ is well defined, i.e., the value of $d$ is unchanged when both $b$ and $q^w$ are multiplied by the
same constant. From the definition of the characteristic function $d$, we see that for any
rational $\frac{a}{q^n}$ with $0<a<q^n$, $d$ is a positive integer. In
each iteration of the algorithm, we hope to reduce $d(p_k)$ such that it can reach $1$ after a certain number of iterations.
If $d=1$, this means the desired probability can be realized using a single pswitch and the construction is done.
During this process, we keep each successive probability
$p_k$ in the form of $\frac{b}{q^w}$, since only this kind of probabilities can be realized with the pswitch set $S$.
Now, we describe the algorithm as follows.

\begin{Algorithm}[Backward algorithm to realize $p_1=\frac{a}{q^n}$ with $0<a<q^n$ and
  pswitch set $S=\{\frac{1}{q},\frac{2}{q},\dots,\frac{q-1}{q}\}$]
\label{alg1}\quad
\begin{enumerate}
  \item Set $k=1$, starting with an empty circuit.
  \item Let \begin{eqnarray*} h(x_k,p_k) = \left\{\begin{array}{cc}
        \frac{p_k}{x_k} & \textrm{if } x_k>p_k \textrm{ (series),}\\
        \frac{p_k-x_k}{1-x_k} & \textrm{if } x_k<p_k \textrm{ (parallel)}. \\
      \end{array}
    \right.\end{eqnarray*}
  We find the optimal $x_{k}\in S$ that minimizes $d(p_{k+1})$ with $p_{k+1}=h(x_k,p_k)$. If $p_{k+1}=\frac{b}{q^w}$, then
  $$d(p_{k+1})=d\left(\frac{b}{q^w}\right)=\frac{q^{w-1}}{\operatorname{gcd}(b,q^{w-1})}.$$
  \item Insert pswitch $x_k$ to the circuit. If $x_k> p_k$, the
    pswitch is inserted in series; otherwise, it is inserted in
    parallel. Then we set $p_{k+1}=h(x_k,p_k).$
  \item Let $k=k+1$.
  \item Repeat steps 2--4 until $p_k$ can be realized using a single
    pswitch. Then insert $p_k$ into the circuit.
\end{enumerate}
\end{Algorithm}

\begin{figure}[!t]
\includegraphics[width=3.6in]{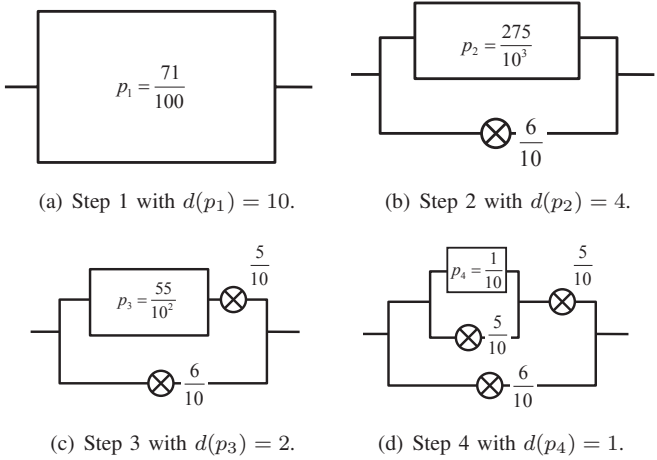}
  \caption{The procedure to realize $\frac{71}{100}$ for a given
    pswitch set $S=\{\frac{1}{10},\frac{2}{10},\dots,\frac{9}{10}\}$.}
\label{fig_alg1}
\end{figure}

In Algorithm \ref{alg1}, the characteristic function $d(p_k)$ strictly
decreases as $k$ increases, until it reaches 1. Finally, $p_k$ can be replaced by a single pswitch and
the construction is done. Fig.~\ref{fig_alg1} gives an example
of a circuit realized by this algorithm. At the beginning, we
have $p_1=\frac{71}{10^2},$ with $d(p_1)=10$. Then we add the ``best''
pswitch to minimize $d(p_2)$, where the optimal pswitch is
$\frac{6}{10}$. Since $\frac{6}{10}<\frac{71}{100}$, we insert the pswitch in
parallel, making $d(p_2)=4$. Repeating this process, we have
$$p_1=\frac{71}{10^2}, \textrm{ } p_2=\frac{275}{10^3}, \textrm{ } p_3=\frac{55}{10^2}, \textrm{ } p_4=\frac{1}{10},$$
with corresponding characteristic functions
$$d(p_1)=10, \textrm{ } d(p_2)=4, \textrm{ } d(p_3)=2, \textrm{ } d(p_4)=1.$$

In the following theorem, we show that if $q$ is a multiple of $2$ or $3$, then Algorithm \ref{alg1} realizes any
rational $\frac{a}{q^n}$ with $0<a<q^n$.

\begin{Theorem}
  Given a pswitch set
  $S=\{\frac{1}{q},\frac{2}{q},\dots,\frac{q-1}{q}\}$, if $q$ is a
  multiple of $2$ or $3$, then Algorithm \ref{alg1} realizes any
  rational $\frac{a}{q^n}$ with $0<a<q^n$, using an ssp circuit with
  a finite number of pswitches.\label{Theorem6}
\end{Theorem}

\proof The characteristic function $d(p_1)$ of the initial probability
$p_1$ is bounded by $q^{n-1}$. We only need to prove that there exists
an integer $m$ such that $d(p_m)=1$, i.e., $p_m$ can be realized by a
single pswitch. Hence the desired probability $p_1$ can be realized by an
ssp circuit with $m$ pswitches. It is enough to show that the
characteristic function $d(p_k)$ decreases as $k$ increases.

\begin{figure}[!t]
\includegraphics[width=3.6in]{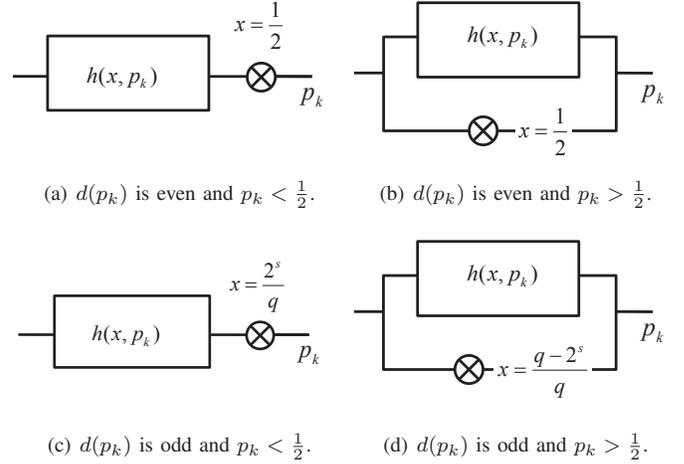}
\caption{When $q$ is even, the way to add a pswitch $x\in S$ such that $d(h(x,p_k))<d(p_k)$.}
\label{fig_insertway}
\end{figure}

First, we consider the case where $q$ is even. We will show that for
any $p_k=\frac{b}{q^w}$, there exists $x\in S$ such that
$d(h(x,p_k))<d(p_k)$. See Fig.~\ref{fig_insertway}, depending on the values of $p_k$ and $d(p_k)$,
we have four different cases of inserting a pswitch $x$ such that $d(h(x,p_k))<d(p_k)$.

\begin{enumerate}
\item If $d(p_k)$ is even and $p_k<\frac{1}{2}$, let $x=\frac{1}{2}$
  and insert the pswitch in series.
\item If $d(p_k)$ is even and $p_k>\frac{1}{2}$, let $x=\frac{1}{2}$
  and insert the pswitch in parallel.
\item If $d(p_k)$ is odd and $p_k<\frac{1}{2}$, let $x=\frac{2^s}{q}$
  with $s=\lfloor\log_2 q\rfloor$ and insert the pswitch in series.
\item If $d(p_k)$ is odd and $p_k>\frac{1}{2}$, let
  $x=\frac{q-2^s}{q}$ with $s=\lfloor\log_2 q\rfloor$ and insert the
  pswitch in parallel.
\end{enumerate}

By checking all the cases to insert a pswitch, it is straightforward to see that when $d(p_k)$ is even,
$d(h(x,p_k))\leq \frac{1}{2}d(p_k)$, and when $d(p_k)$ is odd,
$$d(h(x,p_k))\leq \frac{2^sd(p_k)}{\operatorname{gcd}(q,2^sd(p_k))}<d(p_k).$$
Since $x_k$ is optimal in each step of Algorithm \ref{alg1}, we
have $$d(p_{k+1})=d(h(x_k,p_k))\leq d(h(x,p_k))<d(p_k).$$ Finally, we
can conclude that when $q$ is even, there exists an integer $m$ such
that $d(p_m)=1$. Consequently, $p_1$ can be
realized with at most $m$ pswitches.

Similarly, when $q$ is odd and a multiple of $3$, if
$p_k=\frac{b}{q^w}$, we can always insert a pswitch $x\in S$ such that
$d(h(x,p_k))<d(p_k)$, as follows:
\begin{enumerate}
\item If $d(p_k)\mod 3=0$ and $p_k\leq\frac{1}{3}$, let
  $x=\frac{1}{3}$, and insert the pswitch in series.
\item If $d(p_k)\mod 3=0$ and $\frac{1}{3}<p_k\leq\frac{2}{3}$ with
  even $b$, let $x=\frac{2}{3}$, and insert the pswitch in series.
\item If $d(p_k)\mod 3=0$ and $\frac{1}{3}<p_k\leq\frac{2}{3}$ with
  odd $b$, let $x=\frac{2}{3}$, and insert the pswitch in parallel.
\item If $d(p_k)\mod 3=0$ and $p_k>\frac{2}{3}$, let $x=\frac{2}{3}$,
  and insert the pswitch in parallel.
\item If $d(p_k)\mod 3\neq0$ and $p_k\leq\frac{1}{3}$, let
  $x=\frac{3^s}{q}$ with $s=\lfloor\log_3 q\rfloor$, and insert the
  pswitch in series.
\item If $d(p_k)\mod 3\neq0$ and $\frac{1}{3}<p_k\leq\frac{2}{3}$ with
  even $b$, let $x=\frac{2\cdot3^s}{q}$ with $s=\lfloor\log_3
  q\rfloor$, and insert the pswitch in series.
\item If $d(p_k)\mod 3\neq0$ and $\frac{1}{3}<p_k\leq\frac{2}{3}$ with
  odd $b$, let $x=\frac{q-2\cdot3^s}{q}$ with $s=\lfloor\log_3
  q\rfloor$, and insert the pswitch in parallel.
\item If $d(p_k)\mod 3\neq0$ and $p_k>\frac{2}{3}$, let
  $x=\frac{q-3^s}{q}$ with $s=\lfloor\log_3 q\rfloor$, and insert the
  pswitch in parallel.
\end{enumerate}
Finally, we can conclude that $p_1$ can be
realized with a finite number of pswitches when $q$ is odd and a multiple of $3$.
\hfill\QED

\begin{figure}[!t]
\centering
\includegraphics[width=3.6in]{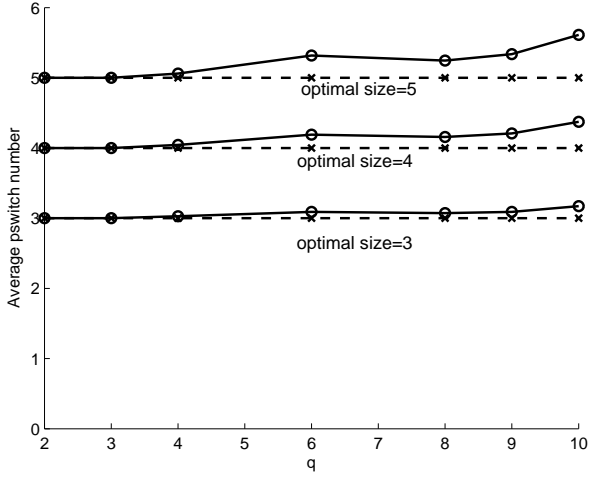}
\caption{For each $q$, the average number of pswitches used in
  Algorithm \ref{alg1} to realize the rational probabilities when their optimal size is
  $n$.} \label{fig_gba_op1}
\end{figure}

For each value $q \in \{2,3,4,6,8,9,10\}$, we enumerate all rational
numbers with optimal size $n\in(3,4,5)$. Here, we say that a desired
probability is realized with optimal size if it cannot be realized
with fewer pswitches. As a comparison, we use Algorithm \ref{alg1} to realize
these rational numbers again. Fig.~\ref{fig_gba_op1} presents the average number
of pswitches required using Algorithm \ref{alg1} when the optimal size is $n$.
It is shown that when $q$ is a multiple of $2$ or $3$, Algorithm \ref{alg1} can construct
circuits with almost optimal size.

The next theorem gives an upper bound for the size of the circuits when $q$ is even.

\begin{Theorem}[Upper bound of circuit size when $q$ is even]
  Suppose $q$ is even. Given a pswitch set
  $S=\{\frac{1}{q},\frac{2}{q},\dots,\frac{q-1}{q}\}$, any rational
  $\frac{a}{q^n}$ with $0<a<q^n$ can be realized by an ssp circuit,
  using at most $\lceil \log_2 q\rceil (n-1)+1$
  pswitches.\label{Theorem1}
\end{Theorem}

\proof In order to achieve this upper bound, we use a modified version
of Algorithm \ref{alg1}. Instead of inserting the optimal pswitch
$x_k$, we insert the pswitch $x$ described in
Fig.~\ref{fig_insertway} as the $k$th pswitch. The resulting characteristic function has
the following properties:

\begin{enumerate}
\item[(1)] $d(p_k)$ decreases as $k$ increases, and when $d(p_m)=1$ for
  some $m$, the procedure stops.
\item[(2)] If $d(p_k)$ is even, then $d(p_{k+1})$ is a factor of
  $\frac{d(p_k)}{2}$.
\item[(3)] If $d(p_k)$ is odd, then $d(p_{k+1})$ is a factor of $\frac{2^s
    d(p_k)}{\operatorname{gcd}(q,2^sd(p_k))}$.
\end{enumerate}

We define
$$N=\min\{k|k\in (1,2,3,\ldots), d(p_k)=1\},$$
then $N$ is the number of required pswitches. We
only need to prove that $N\leq \lceil \log_2 q\rceil (n-1)+1$.
Since $q$ is even, we can write $q=2^c$ or $q=2^ct$, where $t>1$ is
odd.

Let us first consider the case of $q=2^c$. At the beginning, $d(p_1)$
is a factor of $q^{n-1}$, so according to property (2), we can get
$$N\leq c(n-1)+1=\lceil\log_2 q\rceil (n-1)+1.$$

In the case of $q=2^ct$, let us define a set $M$ as
$$M=\{k|k>0, d(p_k) \textrm{ is odd}\},$$
and let $M_i$ be the $i$th smallest element in $M$.
According to properties (2) and (3) and the fact that $d(p_1)$ is a factor of
$q^{n-1}$, we see that $d(p_{M_i})$ is a factor of $q^{n-i}$.
Therefore, there exits a minimal $k$, with $k\leq n$, such that
$d(p_{M_k})=1$. Then $N=M_k$.

Based on properties (2) and (3), we also see that
$$M_1 \leq c(n-1)+1,$$
and
$$M_{i+1}-M_i\leq s-c.$$
Therefore,
\begin{eqnarray*}
N & \leq & \sum_{i=1}^{n-1}(M_{i+1}-M_i)+M_1\leq s(n-1)+1 \\
& = &\lceil\log_2 q\rceil (n-1)+1.
\end{eqnarray*}
This completes the proof. \hfill\QED

Using the similar methods, we can prove the following theorems as well when $q$ is a multiple of $3$ or
$6$. Note that Theorem \ref{Theorem1} also applies to the case that $q$ is a
multiple of $6$, but Theorem \ref{Theorem3} provides a tighter upper bound.

\begin{Theorem}[Upper bound of circuit size when $q$ is odd and a
  multiple of $3$]
  Given a pswitch set
  $S=\{\frac{1}{q},\frac{2}{q},\dots,\frac{q-1}{q}\}$, if $q$ is odd
  and a multiple of $3$, then any rational $\frac{a}{q^n}$ with
  $0<a<q^n$ can be realized using an ssp circuit with at most $\lceil
  \log_3 q\rceil (n-1)+1$ pswitches. \label{Theorem2}
\end{Theorem}

\begin{Theorem}[Upper bound of circuit size when $q$ is a multiple of $6$]\label{Theorem3}
  Given a pswitch set
  $S=\{\frac{1}{q},\frac{2}{q},\dots,\frac{q-1}{q}\}$, if $q$ is
  multiple of 6, all rational $\frac{a}{q^n}$ with $0<a<q^n$ can be
  realized by an ssp circuit with at most $N$ pswitches, where
\begin{eqnarray*} N\leq \left\{\begin{array}{ll}
      (2s)(n-1)+1&(\textrm{if }6^s= q), \\
      (2s+1)(n-1)+1&(\textrm{if }\frac{q}{2}\leq 6^s< q), \\
      (2s+2)(n-1)+1&(\textrm{if }\frac{q}{3}\leq 6^s< \frac{q}{2}),\\
      (2s+3)(n-1)+1&(\textrm{if }\frac{q}{6}< 6^s\leq \frac{q}{3}).\\
           \end{array}
\right.\end{eqnarray*}
\end{Theorem}

\subsection{Prime Number Larger Than $3$}

We proved that if $q$ is a multiple of $2$ or $3$, all rational
$\frac{a}{q^n}$ can be realized with a finite number of pswitches. We want to know whether
this result also holds if $q$ is an arbitrary number greater than 2. Unfortunately,
the answer is negative.

\begin{Lemma}
  Suppose $q$ is a prime number. Given a pswitch set
  $S=\{\frac{1}{q},\frac{2}{q},\dots,\frac{q-1}{q}\}$, if a rational
  $\frac{a}{q^n}$ cannot be realized by an sp circuit with $n$
  pswitches, then it cannot be realized using an sp circuit with any
  number of pswitches.\label{Theorem4}
\end{Lemma}

\proof Assume there exits a rational $\frac{a}{q^n}$ which cannot be
realized by an sp circuit with $n$ pswitches, but can be realized with
at least $l > n$ pswitches. Further, suppose that this $l$ is minimal for all rationals with denominator $q^k$.
Under these assumptions, we will prove that there exists a rational $\frac{a'}{q^{n'}}$ which cannot be realized
with $n'$ pswitches but can be realized with $l'$ pswitches such that $l'<l$. This conclusion contradicts the assumption that $l$ is minimal.

According to the definition of sp circuits, we know that
$\frac{a}{q^n}$ can be realized by connecting two sp circuits $C_1$
and $C_2$ in series or in parallel. Assume $C_1$ consists of $l_1$
pswitches and is closed with probability $\frac{b_1}{q^{l_1}}$, and
$C_2$ consists of $l_2$ pswitches and is closed with probability
$\frac{b_2}{q^{l_2}}$, where $l_1+l_2=l$.

If $C_1$ and $C_2$ are connected in series, we can get
$$\frac{b_1}{q^{l_1}}\cdot\frac{b_2}{q^{l_2}}=\frac{a}{q^n}.$$

Therefore, $b_1b_2=aq^{l-n}$, where $b_1b_2$ is a multiple of
$q$. Since $q$ is a prime number, either $b_1$ or $b_2$ is a multiple
of $q$. Without loss of generality, assume $b_1$ is a multiple of $q$,
and we write $b_1 = cq$. Consider the probability $\frac{c}{q^{l_1-1}}$,
which can be realized with $C_1$, using $l_1$ pswitches. Assume that
the same probability can also be realized with another sp circuit
$C_3$, using $l_1-1$ pswitches. By connecting $C_3$ and $C_2$ in
series, we can realize $\frac{a}{q^n}$ with $l_1-1+l_2=l-1$ pswitches,
contradicting the assumption that $\frac{a}{q^n}$ cannot be realized
with less than $l$ pswitches. Therefore, we see that $\frac{c}{q^{l_1-1}}$ cannot
be realized with $l_1-1$ pswitches, but it can be realized with $l_1$
pswitches. Since $l_1<l$, this also contradicts our assumption that
$l$ is minimal.

If $C_1$ and $C_2$ are connected in parallel, we have
$$\frac{b_1}{q^{l_1}}+\frac{b_2}{q^{l_2}}-\frac{b_1}{q^{l_1}}\cdot\frac{b_2}{q^{l_2}}=\frac{a}{q^{n}}.$$

Therefore, $b_1b_2=b_1 q^{l_2}+b_2q^{l_1}-aq^{l-n}$. Using a similar argument as above,  we can conclude that either $b_1$ or $b_2$ is a multiple of
$q$. Then either (1) $\frac{a}{q^l}$ can be realized with less than
$l$ pswitches or (2) $l$ is not optimal, yielding a
contradiction. This proves the lemma. \hfill\QED

Based on the lemma above, it is easy to get the following theorem.

\begin{Theorem}[When $q$ is a prime number larger than $3$]\label{theorem_switching_5}
  For a prime number $q>3$, there exists an integer $a$, with
  $0<a<q^n$, such that $\frac{a}{q^n}$ cannot be realized using an sp
  circuit whenever $n\geq 2$.
\end{Theorem}

\emph{Proof:} The conclusion follows Lemma \ref{Theorem4} and the following result in \cite{Wilhelm2008}:
For any
$q>3$, no pswitch set containing all $\frac{a}{q}$, with $0<a<q$, can
realize all $P_r(C)=\frac{b}{q^2}$, with $0<b<q^2$, using at most $2$
pswitches. \hfill\QED

\section{Probability Approximation}
\label{section_scenario2}

In this section, we consider a general case where given an arbitrary pswitch set, we want to realize a desired probability. Clearly, not every desired
probability $p_d$ can be realized without any error using a finite number of pswitches for a fixed pswitch set $S$.
So the question is whether we can construct a circuit with
at most $n$ pswitches such that it can approximate the desired probability very well. Namely, the difference between the probability of the constructed circuit and the desired probability should be as small as possible.

\subsection{Greedy Algorithm}

Given an arbitrary pswitch set $S$ with $|S|\geq 2$, it is not easy
to find the optimal circuit (ssp circuit) with $n$ pswitches which approximates the
desired probability $p_d$. As we discussed in the last section, a backward algorithm
provides $|S|$ choices for each successive insertion. To find the optimal circuit, we may have to search through $|S|^n$ different combinations. As $|S|$ or $n$ increases, the number of combinations
will increase dramatically. In order to reduce the search space,
we propose a greedy algorithm: In each step, we insert $m$ pswitches, which are the ``best" locally.
Normally, $m$ is a very small constant. Since each
step has complexity $|S|^m$, the total number of possible combinations is reduced to
$|S|^m\frac{n}{m}$, which is much smaller than $|S|^n$ when $|S|\geq 2$ and $n$ is large.
Now, we describe this greedy algorithm briefly. The same notations $x_1, x_2, \ldots$ and
$p_1, p_2, \ldots$ are used, as those described  for the backward algorithms: $x_k$ indicates
the $k$th pswitch inserted and $p_k$ indicates the desired probability of the subcircuit constructed by $x_k, x_{k+1}, \ldots$

\begin{Algorithm}[Greedy algorithm with step-length $m$]
\label{alg_case3_2}\quad
\begin{enumerate}
\item Assume that the desired probability is $p_1$. Set $k=1$ and start with an empty circuit.
\item Select the optimal $x^m=(x_1,x_2,\dots,x_m)\in S^m$ to minimize
$f(x^m,S,p_k)$, which will be specified later, and this $x^m$ is denoted as $x^*=(x_1^*,x_2^*,\dots,x_m^*)$.
\item Insert $m$ pswitches $x_1^*, x_2^*, \dots, x_m^*$ one by one into the circuit in backward direction. During this process,
calculate $p_{k+1}, p_{k+2}, \ldots, p_{k+m}$ one by one and update $k$ as $k+m$.
\item Repeat steps 2 and 3 for $\lfloor\frac{n}{m}\rfloor$ times.
\item Construct a new circuit with $n-\lfloor\frac{n}{m}\rfloor m$ pswitches such that its probability is closest to $p_k$, then replace $p_k$ with this new circuit.
\end{enumerate}
\end{Algorithm}

So far, according to the backward algorithm described in Section \ref{subsec_backwardalg}, we know
how to finish step 3, including how to insert $m$ pswitches one by one into a circuit in a backward direction, and how to update $p_k$. The only thing unclear in the procedure above is the expression of $f(x^m,S, p_k)$.

In order to get a good expression for $f(x^m, S, p_k)$, we study how errors propagate in a backward algorithm.
Note that in a backward algorithm, we insert pswitches $x_1,x_2, \ldots,x_n$ one by one: if
$x_k>p_k$, then $x_k$ is inserted in series; if $x_k<p_k$, then $x_k$ is
inserted in parallel. Now, given a circuit $C$ with size $n$ constructed using a backward algorithm,
we let $C^{(k)}$ denote the subcircuit constructed by $x_{k_1}, x_{k_1+1}, \ldots, x_n$ and call
$|P(C^{(k)})-p_k|$ as the approximation error of $p_k$, denoted by $e_k$. In the following theorem,
we will show how $e_{k_1}$ affects that of $e_{k_2}$ for $k_2<k_1$ after inserting pswitches
$x_{k_2}, \ldots,x_{k_1-1}$.

\begin{Lemma}
  In a backward algorithm, let $p_k$ denote the desired probability of the subcircuit $C^{(k)}$ constructed by $x_k, x_{k+1},\ldots,x_n$, and let $e_k$ denote the
  approximation error of $p_k$. Then for any $k_2<k_1\leq n$, we have
$$e_{k_2}=\left(\prod_{i=k_1}^{k_2-1} r(x_i)\right)e_{k_1},$$
where
$$r(x_i)=\left\{\begin{array}{cc}
    x_i & \textrm{if $x_i$ is inserted in series,}  \\
    1- x_i & \textrm{if $x_i$ is inserted in parallel.}
                \end{array}
              \right.$$ \label{lemma_error}
\end{Lemma}

\proof We only need to prove that for any $k$ less than the circuit
size, the following result holds:
$$e_{k}=r(x_k)e_{k+1}.$$

When $x_k=p_k$, we have $e_{k}=e_{k+1}=0$, so the result is trivial.

When $x_k>p_k$, then $x_k$ is inserted in series. In this case, we have
$$p_{k+1}x_k=p_k,$$
and
$$P(C^{(k+1)})x_k=P(C^{(k)}).$$
As a result, the approximation error of $p_k$ is
\begin{eqnarray*}
  e_k &=& |P(C^{(k)})-p_k| \\
  &=& |P(C^{(k+1)})x_k-p_{k+1}x_k|\\
  &=& x_k e_{k+1}.
\end{eqnarray*}

When $x_k<p_k$, then $x_k$ is inserted in parallel. In this case, we have
$$p_{k+1}+x_k-p_{k+1}x_k=p_k,$$
and
$$P(C^{(k+1)})+x_k-P(C^{(k+1)})x_k=P(C^{(k)}).$$
As a result, the approximation error of $p_k$ is
\begin{eqnarray*}
  e_k
   &=& |P(C^{(k)})-p_k| \\
  &=& |P(C^{(k+1)})+x_k-P(C^{(k+1)})x_k\\
  &&-(p_{k+1}+x_k-p_{k+1}x_k)|\\
  &=& (1-x_k) e_{k+1}.
\end{eqnarray*}

This completes the proof.
\hfill\QED

In each step of the greedy algorithm, our goal is to minimize $e_k$, the approximation error of $p_k$. According to
the lemma above, we know that
$$e_k=\left(\prod_{i=k}^{k+m-1} r(x_i)\right)e_{k+m},$$
where the term $e_{k+m}$ is unknown. But we can minimize $\prod_{i=k}^{k+m-1} r(x_i)$ such that $e_k$ is as small as possible.

Based on the above discussion, we express $f(x,S,p_k)$ as
\begin{equation}f(x,S,p_k)=\prod_{i=1}^m r(x_i),\label{equation_switching_2}\end{equation}
with
$$r(x_i)=\left\{\begin{array}{cc}
    x_i & \textrm{if $x_i$ is inserted in series,}  \\
    1- x_i & \textrm{if $x_i$ is inserted in parallel.}
                \end{array}
              \right.
$$
In the rest of this section, based on this expression for $f(x,S,p_k)$,
we show that the greedy algorithm has good performance in reducing the approximation
error of $p_d$.

\subsection{Approximation Error when $|S|=1$}

\begin{figure}[!ht]
\centering
\includegraphics[width=1.8in]{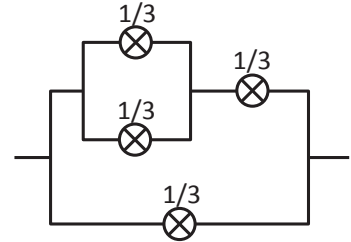}
\caption{This circuit approximates $\frac{1}{2}$ using $4$ pswitches of probability $\frac{1}{3}$.} \label{fig_approximationm1}
\end{figure}

When $S$ has only one element, say $S = \{p\}$, the greedy algorithm above can become really simple.
If $p_k>p_k$, then we insert one pswitch in parallel; otherwise, we insert it in series.
Fig.~\ref{fig_approximationm1} demonstrates how to approximate
$\frac{1}{2}$ using four pswitches with the same probability $\frac{1}{3}$. Initially,
$p_1=\frac{1}{2}>\frac{1}{3}$, so we insert $\frac{1}{3}$ in
parallel. As a result,
$p_2=\frac{\frac{1}{2}-\frac{1}{3}}{1-\frac{1}{3}}=\frac{1}{4}<\frac{1}{3}$,
so we insert the second pswitch in series. The final probability
of the circuit in Fig.~\ref{fig_approximationm1} is $\frac{37}{81}$,
which is close to $\frac{1}{2}$.

Note that in the greedy algorithm, when $p$ is close to $\frac{1}{2}$, the probability of the
resulting circuit will quickly converge to the desired
probability. But when $p$ is close to $0$ or $1$, the convergence speed is slower. In the following theorem, we provide
an upper bound for the approximation error of the desired probability when $|S|=1$.

\begin{Theorem}[Approximation error when $|S|=1$]
  Given $n$ pswitches, each with probability $p$, and a desired
  probability $p_d$, the greedy algorithm (Algorithm \ref{alg_case3_2}) with $m=1$ generates an ssp
  circuit $C$ with approximation error
$$e=|p_d-P(C)|\leq \frac{(\max\{p,1-p\})^n}{2},$$
where equality is achieved when
$$p_d=f_n(p)=\left\{\begin{array}{cc}
    1-\frac{(\max\{p,1-p\})^n}{2} &  \textrm{ if $p < \frac{1}{2}$,}\\
    \frac{(\max\{p,1-p\})^n}{2} & \textrm{ if $p > \frac{1}{2}$.}
             \end{array}\right. $$
\end{Theorem}

\proof In the following proof, we only consider the case when
$p<\frac{1}{2}$. From duality, the result will also hold for
$p>\frac{1}{2}$.

We induct on the number of pswitches. For one pswitch, the result is
trivial: the worst-case desired probability is $p+\frac{1-p}{2}$, with
approximation error $\frac{1-p}{2}$. Now assume the result of the
theorem holds for $n$ pswitches, we want to prove that it also holds
for $n+1$ pswitches.

Let $p_1=p_d$ be approximated with $n+1$ pswitches using Algorithm
\ref{alg_case3_2}. At the beginning, one pswitch is inserted in series
if $p_d<p$, or in parallel if $p_d>p$. According to Lemma
\ref{lemma_error}, we know that the approximation error of $p_1$ is
$$e_1=r(p)e_2,$$
where $r(p)\leq \max\{p,1-p\}$, and $e_2$ is the approximation error
of $p_2$. According to our assumption, we know that
$$e_2\leq \frac{(\max\{p,1-p\})^n}{2}.$$

So we have
$$e_1\leq \frac{(\max\{p,1-p\})^{n+1}}{2}.$$

Note that equality is achieved if $r(p)=\max\{p,1-p\}$ and
$e_2=\frac{(\max\{p,1-p\})^n}{2}$. In this case,
$p_2=f_n(p)\geq \frac{1}{2}> p$ and the last pswitch is inserted in parallel.
As a result, we have
$$f_{n+1}(p)=f_n(p)+p-f_n(p)p=1-\frac{(1-p)^{n+1}}{2}$$
as described in the theorem. This completes the proof. \hfill\QED

If we let $p=\frac{1}{2}$, the theorem shows that for any
desired probability $p_d$ and any integer $n$, we can find an ssp
circuit with $n$ pswitches to approximate $p_d$, such that the
approximation error is at most $\frac{1}{2 q^n}$.  This agrees with the result in \cite{Wilhelm2008}: Given a pswitch set $S=\{\frac{1}{2}\}$, all rational
$\frac{a}{2^n}$, with $0<a<q^n,$ can be realized using at most $n$
pswitches.

\subsection{Approximation Error when $|S|>1$}

In this subsection, we show that using the greedy algorithm (Algorithm \ref{alg_case3_2})
with small $m$, such as $1$ or $2$, we can construct a circuit to obtain a
good approximation of any desired probability. Here, given a pswitch set $S=\{s_1,s_2,\ldots,s_{|S|}\}$,
we define its maximal interval $\Delta$ as
$$\Delta=\max_{i=0}^{|S|}|s_{i+1}-s_{i}|,$$
where we let $s_0=0$ and $s_{|S|+1}=1$. In the following theorems,
we will see that the approximation error of the greedy algorithm depends on $\Delta$, and can decrease
rapidly as $n$ increases.

Let us first consider the case $m=1$:
\begin{Theorem}[Approximation error for $m=1$]
Assume we have the pswitch set $S=\{s_1,s_2,\dots,s_{|S|}\}$ with maximal interval $\Delta$.
For any desired probability $p_d$ and any integer $n$, Algorithm
\ref{alg_case3_2} with $m=1$ yields an ssp circuit with at most
$n$ pswitches, such that the approximation error $e$ satisfies
$$e\leq \frac{\Delta}{2}\left(\frac{(3+\Delta)\Delta}{2}\right)^{\lceil\frac{n}{2}\rceil-1}.$$
\end{Theorem}

\proof In the following proof, we only consider the case that $n$ is
odd. If the result holds for odd $n$, then the result
will also hold for even $n$. In order to simplify the proof, we
assume that $s_0=0$ and $s_{|S|+1}=1$ also belong to $S$; i.e.,
there are pswitches with probability $0$ or $1$. This assumption will not affect our conclusion.

We write $n=2k+1$ and induction on $k$. When $k=0$, the result is trivial, since the approximation error of one pswitch satisfies $e\leq \frac{\Delta}{2}$. Assume the
result holds for $2k+1$ pswitches. We want to show that the result
also holds for $2(k+1)+1$ pswitches.

When $m=1$ in the greedy algorithm, if we want to approximate
$p_1=p_d$ with $2(k+1)+1$ pswitches, we should insert a pswitch with
probability $\arg\min_{x}f(x, S,p_1)$ in the first step, where $f(x,S, p_1)$ is defined in (\ref{equation_switching_2}).

Let $x_{\mathrm{upper}}=\min\{x\in S| x>p_1 \}$ and $x_{\mathrm{lower}}=\max\{x\in
S|x<p_1\}$. Since $0\in S$ and $1\in S$, we know that
$x_{\mathrm{upper}}$ and $x_{\mathrm{lower}}$ exist.

(1) We first consider the case that $1-x_{\mathrm{lower}}\leq x_{\mathrm{upper}}$. In
this case, we insert $x_{\mathrm{lower}}$ in parallel as the first pswitch. Therefore, we can get
$$p_2=\frac{p_1-x_{\mathrm{lower}}}{1-x_{\mathrm{lower}}}.$$

According to the definition of $\Delta$, there exists a pswitch
$x\in S$ such that $p_2\leq x < p_2+\Delta$. Assume in the algorithm, we insert
pswitch $x^*$ as the second one. Since $x^*$ is locally optimal, we have
$$f(x^*, S, p_2) \leq f(x, S, p_2) < p_2+ \Delta.$$

Assume the approximation error of $p_3$ is $e_3$. According
to Lemma \ref{lemma_error}, we know that the approximation error of
$p_1=p_d$ is
\begin{eqnarray*}
  e_1 &\leq & (p_2+\Delta)(1-x_{\mathrm{lower}})e_3 \\
  &=&  (\frac{p_1-x_{\mathrm{lower}}}{1-x_{\mathrm{lower}}}+\Delta)(1-x_{\mathrm{lower}})e_3 \\
  &=& ((p_1-x_{\mathrm{lower}})+\Delta(1-x_{\mathrm{lower}}) )e_3\\
  &\leq & \Delta(2-x_{\mathrm{lower}})e_3\\
  &\leq & \frac{\Delta(3+x_{\mathrm{upper}}-x_{\mathrm{lower}})}{2}e_3\\
  &\leq &  \frac{\Delta(3+\Delta)}{2}e_3.
\end{eqnarray*}

According to our assumption,
$$e_3\leq \frac{\Delta}{2}(\frac{(3+\Delta)\Delta}{2})^{k}.$$
So
$$e_1\leq \frac{\Delta}{2}(\frac{(3+\Delta)\Delta}{2})^{k+1}.$$
This completes the induction.

(2) When $1-x_{\mathrm{lower}}> x_{\mathrm{upper}}$, we insert $x_{\mathrm{upper}}$ in
series as the first pswitch. Using a similar argument as above, we can also prove that
$$e_1\leq \frac{\Delta}{2}(\frac{(3+\Delta)\Delta}{2})^{k+1}.$$
This completes the proof.  \hfill\QED

In the next theorem, we show that if we increase $m$ from $1$ to $2$,
the upper bound of the approximation error can be reduced furthermore.

\begin{Theorem}[Approximation error for $m=2$]\label{Theorem_m2}
Assume we have the pswitch set $S=\{s_1,s_2,\dots,s_{|S|}\}$ with maximal interval $\Delta$.
For any desired probability $p_d$ and any integer $n$, Algorithm
\ref{alg_case3_2} with $m=2$ yields an ssp circuit with at most
$n$ pswitches, such that the approximation error $e$ satisfies
$$e\leq \frac{\Delta}{2}\left(\frac{(2+\Delta)\Delta}{2}\right)^{\lceil\frac{n}{2}\rceil-1}.$$
\end{Theorem}

\proof
As in the proof for $m=1$, we only consider the case when $n$
is odd, so $n=2k+1$.  In the proof, we use the same notations as those in the case of $m=1$, and assume
$S$ includes $0$ and $1$.

Now we induct on $k$. When $k=0$, the result of the theorem is
trivial. Assume the result holds for $2k+1$ pswitches; we want to
prove that it also holds for $2(k+1)+1$ pswitches. Let $x_{\mathrm{upper}}=\min\{x\in S| x>p_1 \}$ and $x_{\mathrm{lower}}=\max\{x\in
S|x<p_1\}$, we will consider two different cases as follows.

(1) If $p_1\leq \frac{x_{\mathrm{upper}}+x_{\mathrm{lower}}+\Delta(x_{\mathrm{upper}}+x_{\mathrm{lower}}-1)}{2}$,
we consider the following way to insert two pswitches: First insert $x_1=x_{\mathrm{lower}}$ in parallel, and we get
$$p_2=\frac{p_1-x_{\mathrm{lower}}}{1-x_{\mathrm{lower}}}.$$
There exists a pswitch $x_2\in S$ such that $p_2\leq x_2 <
p_2+\Delta$. Then we insert $x_2$ in series as the second pswitch. In this case, letting $x=(x_1,x_2)$,
we have
$$f(x,S,p_1)\leq (p_2+\Delta) (1- x_{\mathrm{lower}}).$$

Let $x^*=(x_1^*, x_2^*)$ be the two pswitches inserted by the algorithm with $m=2$, then the approximation error of $p_1=p_d$ is
\begin{eqnarray*}
  e_1 &= & f(x^*,S,p_1)e_3 \leq  f(x,S,p_1)e_3 \\
  &\leq & (p_2+\Delta)(1-x_{\mathrm{lower}})e_3 \\
  &=&  \left(\frac{p_1-x_{\mathrm{lower}}}{1-x_{\mathrm{lower}}}+\Delta\right)(1-x_{\mathrm{lower}})e_3.
\end{eqnarray*}

Since $p_1\leq
\frac{x_{\mathrm{upper}}+x_{\mathrm{lower}}+\Delta(x_{\mathrm{upper}}+x_{\mathrm{lower}}-1)}{2}$, we have
\begin{eqnarray*}
  e_1 &\leq & \frac{(x_{\mathrm{upper}}-x_{\mathrm{lower}})(1+\Delta)+\Delta}{2}e_3\\
  &\leq& \frac{\Delta(2+\Delta)}{2}e_3.
\end{eqnarray*}

According to our assumption, we have
$e_3\leq \frac{\Delta}{2}\left(\frac{(2+\Delta)\Delta}{2}\right)^{k}$, so $$e_1\leq \frac{\Delta}{2}\left(\frac{(2+\Delta)\Delta}{2}\right)^{k+1}.$$

This completes the induction.

(2) If $p_1>\frac{x_{\mathrm{upper}}+x_{\mathrm{lower}}+\Delta(x_{\mathrm{upper}}+x_{\mathrm{lower}}-1)}{2}$,
we consider the following way to insert two pswitches: First insert $x_1=x_{\mathrm{upper}}$ in series, and we get
$$p_2=\frac{p_1}{x_{\mathrm{upper}}}.$$
There exists a pswitch $x_2\in S$ such that $p_2-\Delta \leq x_2 <
p_2$. Then we insert $x_2$ in parallel as the second pswitch. In this case, letting $x=(x_1,x_2)$,
we have
$$f(x,S,p_1)\leq(1-(p_2-\Delta)) x_{\mathrm{upper}}.$$

Let $x^*=(x_1^*, x_2^*)$ be the two pswitches inserted by the algorithm with $m=2$, then the approximation error of $p_1=p_d$ is
\begin{eqnarray*}
  e_1 &=& f(x_*,S,p_1)e_3 \\
  &\leq & f(x,S,p_1)e_3 \\
  &\leq & (1-(p_2-\Delta)) x_{\mathrm{upper}} e_3\\
  &=& \left(\frac{x_{\mathrm{upper}}-p_1}{x_{\mathrm{upper}}}+\Delta\right)x_{\mathrm{upper}}e_3.
\end{eqnarray*}
Since
$p_1>\frac{x_{\mathrm{upper}}+x_{\mathrm{lower}}+\Delta(x_{\mathrm{upper}}+x_{\mathrm{lower}}-1)}{2}$, we
have
\begin{eqnarray*}
  e_1 &\leq & \frac{(x_{\mathrm{upper}}-x_{\mathrm{lower}})(1+\Delta)+\Delta}{2}e_3\\
  &\leq& \frac{\Delta(2+\Delta)}{2}e_3.
\end{eqnarray*}

Then we have the same result as the first case. \hfill\QED

According to the two theorems above, when we let $\Delta\rightarrow 0$, the approximation error for $m=1$
 is upper bounded by $\frac{\Delta}{2}\left(\frac{3\Delta}{2}\right)^k$ where $k=\lceil\frac{n}{2}\rceil-1$; and the approximation error
 for $m=2$ is upper bounded by $\frac{\Delta}{2}\cdot\Delta^k$. It shows that the greedy algorithm has good performance
 in terms of approximation error, even when $m$ is very small. Comparing with the case of $m=1$, if we choose $m=2$,
 the probability of the constructed circuit can converge to the desired probability faster as the circuit size $n$
 increases.

In the following theorem, we consider the special case $S=\{\frac{1}{q},\frac{2}{q},\dots,\frac{q-1}{q}\}$ for some integer
$q$. In this case, we obtain a new upper bound for the approximation
error when using the greedy algorithm with $m=2$. This bound is slightly tighter than the one obtained in Theorem \ref{Theorem_m2}.

\begin{Theorem}
  Suppose $S=\{\frac{1}{q},\frac{2}{q},\dots,\frac{q-1}{q}\}$ for
  some integer $q$, with $\Delta=\frac{1}{q}$. For any desired
  probability $p_d$ and any integer $n$, Algorithm \ref{alg_case3_2}
  with $m=2$ constructs an ssp circuit with at most $n$ pswitches such that its
  approximation error
$$e\leq \frac{\Delta}{2}\left(\Delta(1-\Delta)\right)^{\lceil\frac{n}{2}\rceil-1}.$$
\end{Theorem}

\proof The proof is similar to the proof of Theorem \ref{Theorem_m2},
so we simply provide a sketch. Assume that in each step, we insert
two pswitches in the following way (see Fig.~\ref{fig_approximateq}):

\begin{figure}[!ht]
\includegraphics[width=3.6in]{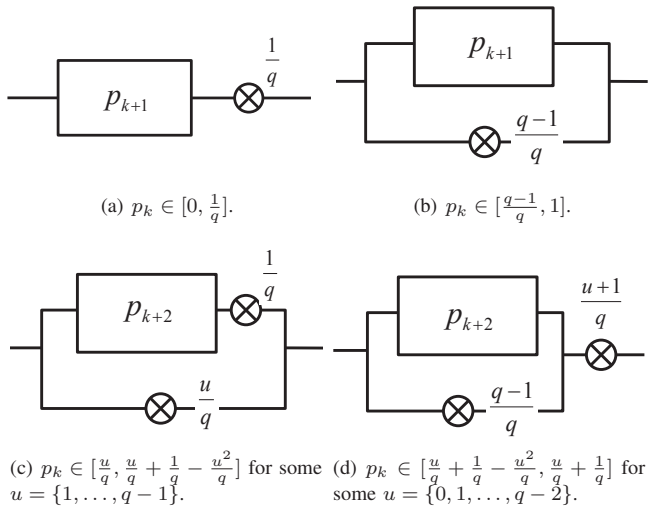}
\caption{Inserting pswitches for different values of $p_k$.}
\label{fig_approximateq}
\end{figure}

(1) If $p_k\in [0,\frac{1}{q}]$, we insert a pswitch $x_1=\frac{1}{q}$
in series, and then insert a pswitch $x_2=\frac{1}{q}$ in series or in
parallel. In this case,
$$f\left(\left(\frac{1}{q},\frac{1}{q}\right),S,p_k\right)\leq \frac{1}{q}\left(1-\frac{1}{q}\right)=\Delta(1-\Delta).$$

(2) If $p_k\in [\frac{q-1}{q},1]$, we insert a pswitch
$x_1=\frac{q-1}{q}$ in parallel, and then insert a pswitch
$x_2=\frac{q-1}{q}$ in series or in parallel. In this case,
$$f\left(\left(\frac{1}{q},\frac{1}{q}\right),S,p_k\right)\leq \frac{1}{q}\left(1-\frac{1}{q}\right)=\Delta(1-\Delta).$$

(3) If $p_k\in [\frac{u}{q},\frac{u}{q}+\frac{1}{q}-\frac{u^2}{q}]$
for some $u=\{1,\dots,q-1\}$, we insert a pswitch $x_1=\frac{u}{q}$ in
parallel, and then insert a pswitch $x_2=\frac{1}{q}$ in series. In
this case,
$$f\left(\left(\frac{u}{q},\frac{1}{q}\right),S,p_k\right)\leq \left(1-\frac{u}{q}\right)\frac{1}{q}\leq\Delta(1-\Delta).$$

(4) If $p_k\in
[\frac{u}{q}+\frac{1}{q}-\frac{u^2}{q},\frac{u}{q}+\frac{1}{q}]$ for
some $u=\{0,1,\dots,q-2\}$, we insert a pswitch $x_1=\frac{u+1}{q}$ in
series, and then insert a pswitch $x_2=\frac{q-1}{q}$ in parallel. In
this case,
\begin{eqnarray*}
f\left(\left(\frac{u+1}{q},\frac{q-1}{q}\right),S,p_k\right) & \leq & \frac{u+1}{q}\left(1-\frac{q-1}{q}\right) \\
& \leq & \Delta(1-\Delta).
\end{eqnarray*}

Based on the above analysis, we know that for any $p_k \in (0,1)$, we
can always find $x=(x_1,x_2)$ such that
$$f((x_1,x_2),S,p_k))\leq \Delta(1-\Delta).$$
Hence, the result of the theorem can be proved by induction.

This completes the proof.\hfill
\QED

\begin{figure}[!ht]
\centering
\includegraphics[width=2.0in]{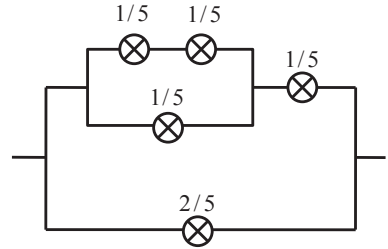}
\caption{The circuit approximates $\frac{3}{7}$ with $5$ pswitches
  from the pswitch set
  $S=\{\frac{1}{5},\frac{2}{5},\dots,\frac{4}{5}\}$.} \label{fig_approximation}
\end{figure}

Fig.~\ref{fig_approximation} shows an example for demonstration.
Assume
$S=\{\frac{1}{5},\frac{2}{5},\frac{3}{5},\frac{4}{5}\}$, and suppose we want
to realize $\frac{3}{7}$ using five pswitches. Using the greedy algorithm with $m=2$, we can get the circuit in Fig.~\ref{fig_approximation},
whose probability is $0.4278$, and approximation error is
$$e=\left |\frac{3}{7}-0.4278\right |=7.3\times10^{-4},$$
which is very small.

\section{Conclusion}
\label{switch_section_conclusion}

In this paper, we have studied the robustness and synthesis of
stochastic switching circuits. We have shown that ssp circuits are
robust against small error perturbations, while general sp circuits
are not. As a result, we focused on constructing ssp circuits to
synthesize or approximate probabilities. We generalized the
results in \cite{Wilhelm2008} and proved that when $q$ is a multiple of
$2$ or $3$, all rational fractions $\frac{a}{q^n}$ can be realized
using ssp circuits when the pswitch set $S =
\{\frac{1}{q},\frac{2}{q},\dots,\frac{q-1}{q}\}$. However, this
property does not hold when $q$ is a prime number greater than $3$.
For a more general case of an arbitrary pswitch set, we proposed a greedy algorithm to construct ssp circuits. This method can approximate any desired probability with low circuit complexity and small errors.

\ifCLASSOPTIONcaptionsoff
  \newpage
\fi

%

%
%
%




\end{document}